\documentclass{aa}

\usepackage{txfonts, graphicx}
\usepackage{natbib}
\bibpunct{(}{)}{;}{a}{}{,}

\newcommand{\tsun}  {${T}_{\odot}$}
\newcommand{\rsun}  {${R}_{\odot}$}

\newcommand{\tefts}{$T_{\rm eff}$~}
\newcommand{\teft}{$T_{\rm eff}$}

\newcommand{\tef}{T_{\rm eff}}
\newcommand{\bc}{$BC$~}
\newcommand{\bC}{$BC$}
\newcommand{\logG}{$\log g$}
\newcommand{\logg}{$\log g$~}
\newcommand{\feh}{$[m/H]$~}
\newcommand{\FEH}{$[m/H]$}
\newcommand{\uvbys}{$uvby-\beta$ }

\newcommand{\MASS} {{2MASS }}

\begin{document}
\title{Effective temperature scale and bolometric corrections from 2MASS 
photometry\thanks {Table \ref{tabsample} is only available in electronic 
form at the CDS via anonymous ftp to {\tt cdsarc.u-strasbg.fr (130.79.128.5)} 
or via {\tt http://cdsweb.u-strasbg.fr/cgi-bin/qcat?]/A+A/}}}
\author{E. Masana \inst{1,2} \and C. Jordi \inst{1,2,3} \and I. Ribas \inst{3,4}}
\institute{Departament d'Astronomia i Meteorologia, Universitat de Barcelona, Avda. Diagonal 647, 08028 Barcelona, Spain 
\and CER for Astrophysics, Particle Physics and Cosmology, associated with Institut de Ci\`encies de l'Espai-CSIC
\and Institut d'Estudis de Espacials de Catalunya (IEEC). Edif. Nexus, C/Gran Capit\`a 2-4, 08034 Barcelona, Spain
\and Institut de Ci\`encies de l'Espai, CSIC, Campus UAB, Fac. Ci\`encies, Torre C5 parell, 2$^{\rm a}$ planta, 08193 Bellaterra, Spain}
\date{Received / Accepted }

\abstract{This paper presents a method to determine effective
temperatures, angular semi-diameters and bolometric corrections for
population I and II FGK type stars based on $V$ and \MASS IR photometry.
Accurate calibration is accomplished by using a sample of solar analogues,
whose average temperature is assumed to be equal to the solar effective
temperature of 5777 K. By taking into account all possible sources of
error we estimate associated uncertainties better than 1\% in effective
temperature and in the range 1.0--2.5\% in angular semi-diameter for
unreddened stars. Comparison of our new temperatures with other
determinations extracted from the literature indicates, in general,
remarkably good agreement. These results suggest that the effective
temperaure scale of FGK stars is currently established with an accuracy
better than 0.5\%--1\%. The application of the method to a sample of
10999 dwarfs in the Hipparcos catalogue allows us to define temperature
and bolometric correction ($K$ band) calibrations as a
function of $(V-K)$, $[m/H]$ and \logg. Bolometric corrections in
the $V$ and $K$ bands as a function of \teft, $[m/H]$ 
and \logg are also given. We provide effective temperatures, angular
semi-diameters, radii and bolometric corrections in the $V$ and
$K$ bands for the 10999 FGK stars in our sample with the
corresponding uncertainties.

 }

\titlerunning{Effective temperature scale and bolometric corrections}
\maketitle

\section{Introduction}

Effective temperature and luminosity are two fundamental stellar
parameters that are crucial to carry out tests of theoretical models of
stellar structure and evolution by comparing them with observations. The
accuracy in the determination of other stellar properties, such as
metallicity, age or radius, hinges on our ability to estimate the
effective temperatures and luminosities.

There are several approaches in the literature to compute effective
temperature and/or luminosity. Except when applied to the Sun, very few of
them are {\it direct} methods that permit an empirical measurement of
these parameters. Usually, {\it semi-empirical} or {\it indirect} methods
are based to a certain extent on stellar atmosphere models. Among the {\it
direct} approaches we find the remarkable work by \citet{code76}, which is
based on interferometric measurements of stellar angular semi-diameters
($\theta$) and total fluxes ($F_{\rm bol}$) at Earth, and the more
recent works of \citet{mozurkewich03} and \citet{kervella04}, also based
on interferometry. On the other hand, {\it indirect} methods are mainly
based on the use of photometry, spectroscopy, or a combination of both. In
the case of the temperatures, although many of the published calibrations
claim to have uncertainties of the order of several tens of degrees,
values obtained by different authors can easily have discrepancies of 100
K or even larger. The reason for such nagging differences must be found
somewhere in the {\it ingredients} of the methods: atmosphere models,
absolute flux calibrations, oscillator strengths, calibration stars, etc.
 
In this paper we present a {\it semi-empirical} method to determine
effective temperatures (\teft) and bolometric corrections ($BC$) from
2MASS\footnote{\tt http://www.ipac.caltech.edu/2mass} $JHK$\footnote
{Throughout the paper, $K$ refers to $K_s$ band.} photometry
\citep{cutri03b} that is applicable to FGK type stars. As all others, our
method is susceptible to problems derived from the uncertainties in the
ingredients mentioned above. However, our approach benefits from two
major features: First, it provides a way to evaluate realistic individual
uncertainties in \teft, $\theta$ and luminosity by considering all the
involved errors; and second, as it is calibrated to use the \MASS
photometry, it allows the calculation of consistent and homogeneous \tefts
and $BC$ for several million stars in the \MASS catalogue. This paper also
provides \teft, angular semi-diameters, radii and $BC$s for 10999 dwarfs
and subdwarfs in the Hipparcos catalogue \citet{ESA97}. Such large sample
has allowed us to construct simple parametric calibrations as a function
of $(V-K)_0$, \feh and \logG. Note that a preliminary version of the method
presented here was already successfully applied to the characterization of
the properties of planet-hosting stars \citep{iribas03a}.

The present paper is organized as follows. Section \ref{sedf} presents the
method and explains in detail the procedure to obtain \tefts and angular
semi-diameters, including the fitting algorithm, zero point corrections
and error estimates. The comparison of our temperatures with several
previous works, both based on photometric and spectroscopic techniques, is
described in Sect. \ref{comparacions}. In Sect. \ref{calib} we present
simple parametric calibrations of \tefts and $BC$ as a function of
$(V-K)_0$, $[m/H]$ and \logg valid for dwarf and subdwarf stars.  
The sample of 10999 stars used to build the calibrations is also described
in this section together with a detailed explanation of the different
contributors to the final uncertainties.  Finally, the results are
discussed in Sect. \ref{disc} and the conclusions of the present work are
presented in Sect. \ref{conclu}.

\section{The Spectral Energy Distribution Fit (SEDF) method}
\label{sedf}

The use of infrared (IR) photometry to determine effective temperatures
was initially proposed by \citet{blackwell77}. Their so-called Infrared
Flux Method (IRFM) uses the ratio between the bolometric flux of the star
and the monochromatic flux at a given infrared wavelength, both measured
at Earth, as the observable quantity. This ratio is then compared with a
theoretical estimate derived from stellar atmosphere models to carry out
the determination of the effective temperature. The IRFM has been widely
used by a number of authors, being most noteworthy the work by
\citet{alonso95, alonso96a, alonso96b}.

The Spectral Energy Distribution Fit (SEDF) method that we propose here
follows a somewhat different approach, namely the fit of the stellar
spectral energy distribution from the optical ($V$) to the IR ($JHK$)
using synthetic photometry computed from stellar atmosphere models. 
Unlike the \cite{alonso96a} implementation of the IRFM, 
which averages temperatures derived individually for each
IR band, our method takes into account the four bands simultaneously (and
naturally). In addition, and also unlike the IRFM, the bolometric flux is
not required {\it a priori} by the SEDF method but results
self-consistently with the temperature. The fitting algorithm (see Sect.
\ref{fit}) minimizes the difference between observed and synthetic
photometry by tuning the values of the effective temperature and the
angular semi-diameter. The $BC$ can be obtained from these two parameters,
and then, when the distance to the star is known, the luminosity is
computed from the $BC$ and the absolute magnitude in a given photometric
band. The uncertainties of the derived parameters (\tefts, angular
semi-diameter and $BC$) are estimated from the errors in the observed and
synthetic photometry as well as in the assumed \FEH, \logg and $A_V$.

From a theoretical point of view, the SEDF method could be applied to
stars of any spectral type and luminosity class. However, the IR flux
becomes very sensitive to metallicity and surface gravity for stars hotter
than 8000 K so that small uncertainties in these parameters translate into
large uncertainties in the effective temperature. In such situation the SEDF
approach becomes inadequate. At the cold end, the accuracy of stellar
atmosphere models limits the use of the method to stars hotter than 4000 K
(molecular opacity plays an important role below this temperature). These
limitations restrict the applicability of the SEDF to FGK type stars.
Fortunately, these stars are very common in the Galaxy and dominate the
content of most of survey catalogues. They are crucial for several key
astrophysical topics, such as the study of the structure and evolution of
the Galaxy, both the disk and the halo, and the characterization of
planet-hosting stars, among others.

\subsection{Calculation of synthetic photometry}
\label{synPhot}

The calculation of the synthetic photometry requires a well-characterized
photometric system, an accurate flux calibration and suitable synthetic
spectra. The work by \citet{cohen03a, cohen03b} provides consistent
absolute flux calibrations in both the visible ($V$) (Landolt system) and
IR (2MASS $JHK$) bands. The calibration given by \citeauthor{cohen03a} is
computed from a set of calibrated templates, using the synthetic Kurucz
spectrum of Vega of \citet{cohen92}. In the case of the IR photometry,
they consider the transmission of the camera and filters, the detector
properties and the Earth's atmosphere characteristics. From the comparison
between observed and synthetic photometry for a set of 9 A-type stars and
24 cool giants, the authors infer the need to introduce a zero point
offset in the synthetic photometry to match the observed \MASS photometry:
0.001$\pm$0.005 mag ($J$); $-0.019\pm$0.007 mag ($H$); 0.017$\pm$0.005 mag
($K$). The calculation of such values is not exempt of some difficulty
since the dispersions of the differences between both photometries
(synthetic and observed) are of the same magnitude as the zero point
offsets themselves.

To compute the syntheric magnitudes we made use of the no-overshoot
Kurucz atmosphere models grid \citep{kurucz79} taken from {\tt
http://kurucz.harvard.edu/grids.html}:
\begin{equation}
m_{\rm syn}^i(\tef, \log g, [m/H])= 2.5 \log \left( {{F_{\rm cal}^i} 
\over {F_i (\tef, \log g, [m/H])}}\right)
\label {eqmag}
\end{equation}
\noindent where $F_{\rm cal}^i$ is the absolute flux calibration given by
\citet{cohen03b} (for $m^i_{\rm cal}=0$) and ${F_i (\tef, g, [m/H])}$ is
the flux in the $i$ band computed from the integration of the model
atmosphere convolved with the transmission function (filter, detector and
Earth's atmosphere) from \citet{cohen03b}:
\begin{equation}
F_i(\tef, \log g, [m/H])= \displaystyle\int_{0}^{\infty} \phi 
(\tef, \log g, [m/H], \lambda){\cal{T}}_i(\lambda) d\lambda 
\label {eqfl}
\end{equation}
\noindent where $\phi (\tef, \log g, [M/H], \lambda)$ is the flux given by
the stellar atmosphere model and ${\cal{T}}_i(\lambda)$ the effective
transmission function in the $i$ band normalized to a peak value of
unity.

\subsection{Fitting algorithm}
\label{fit}

The fitting algorithm is based on the minimization of the $\chi^2$
function defined from the differences between observed (corrected for
interstellar extinction) and synthetic $VJHK$ magnitudes, weighted with
the corresponding error:
\begin{eqnarray}
\chi^2 & = & \left({ {V - A_V - V_{\rm syn}}\over \sigma_V }\right)^2 + 
\left({ {J -A_J- J_{\rm syn}}\over \sigma_J }\right)^2 + \nonumber \\
& + & \left({ {H - A_H - H_{\rm syn}}\over \sigma_H }\right)^2 + 
\left({ {K - A_K - K_{\rm syn}}\over \sigma_K} \right)^2
\label {eqChi}
\end{eqnarray}
This function depends (via the synthetic photometry) on \teft, \logG, \feh
and a magnitude difference $\cal{A}$, which is the ratio between the
synthetic (star's surface) and the observed flux (at Earth)  
(${\cal{A}}=-2.5\log F_{star}/F_{Earth}$). $\cal{A}$ is directly related
to the angular semi-diameter by the following expression:
\begin{equation}
\theta=10^{-0.2\cal{A}}
\label{semidia}
\end{equation}
Although the synthetic photometry depends implicitly on gravity and
metallicity, in practice, the spectral energy distribution in the
optical/IR for our range of temperatures is only {\it weakly dependent} on
these quantities. This fact makes it possible to obtain accurate
temperatures even for stars with poor determinations of \logg and \FEH.

As can be seen, the $\chi^2$ function depends also on the interstellar
absorption $A_V$ (the absorption in the other bands can be computed using
the extinction law of \citet {landolt82}: $A_J = 0.30 A_V$, $A_H = 0.24
A_V$ and $A_K = 0.15 A_V$). In principle, it is possible to consider $A_V$
as a free parameter. However, the strong correlation between \tefts and
$A_V$, especially for the hotter stars, decreases the precision in the
determination of both parameters, with resulting typical uncertainties of
4\% in \tefts and 0.25 mag in $A_V$. Thus, for best performance, $A_V$
should only be considered as a free parameter when its value is suspected
large and no other method for its estimation is available. In general, the
best approach is to fix the value of $A_V$ in Eq. (\ref{eqChi}) from the
estimation by photometric calibrations, for instance.

Therefore, the only two adjustable parameters by the SEDF method in the
present work are \tefts and $\cal{A}$, whereas \logG, \feh and $A_V$ are
fixed parameters. To minimize Eq. (\ref{eqChi}) with respect to these two
parameters we use the Levenberg-Marquardt algorithm \citep{press92}, which
is designed to fit a set of data to a non-linear model. In all our tests,
convergence towards the minimum value of $\chi^2$ was reached rapidly and
unequivocally.

\subsection{Calibration of the SEDF method using solar analogues}
\label{analegs}

The standard procedure for the calibration of an {\it indirect} method to
determine effective temperatures is based on the comparison of the results
with accurate temperatures from {\it direct} methods for a set of stars.
In this way, the list of stars with empirical effective temperatures and
angular semi-diameters from \citet{code76} has been widely used for
calibration purposes. This list has been recently increased with the
the works of \citet{mozurkewich03} and \citet{kervella04}.
Other authors use well-studied stars, such as the
Sun, Vega or Arcturus, to calibrate their methods.

Unfortunately, the few stars with empirical values of \tefts are too
bright to have accurate \MASS photometry and they are of no use to
calibrate the SEDF method. As an alternative, we have used the list of
photometric solar analogues compiled by \citet{cayrel96}. We assume
that, as an ensemble, the average of the effective temperatures of these
photometric solar analogues should be equal to the effective temperature
of the Sun (i.e., 5777~K).

After selecting a subsample of 50 unreddened stars with non-saturated
\MASS photometry from table 1 of \citet{cayrel96}, we computed their
temperatures using the SEDF method. We obtained an average temperature of
$5832\pm14$~K, i.e., 55~K (or $\sim$1\%) higher than the solar effective
temperature. Exactly the same value is obtained if we use the subset of
solar ``effective temperature analogues'' from table 5 of
\citeauthor{cayrel96}. Without a profound analysis of all the ingredients
involved -- from the stellar atmosphere model to the absolute flux
calibration, -- it is very difficult to assess the reasons for such
difference. However, it seems clear that the temperature scale as obtained
from the synthetic photometry alone needs a correction to agree with the
average of the solar analogues. From a formal point of view, this
correction can be computed from the synthetic photometry that results from
forcing a value of \tefts = 5777~K to the entire sample. After doing so,
we replaced the zero points given by \citet{cohen03b} (see Sect.
\ref{synPhot}) by the average difference (for each band) between the
observed and synthetic photometry computed for the solar analogues.
Assuming that there is no offset on the $V$ band, the offsets for the
other bands are: 0.027$\pm$0.003 mag ($J$); 0.075$\pm$0.005 mag ($H$);
0.022$\pm$0.005 mag ($K$). It is interesting to note that both in the case
of \citet{cohen03a} and in our case, the value of the offset in the $H$
band differs significantly from the offsets in $J$ and $K$. 
It should be
stressed that the effective temperatures given by \citet{cayrel96} have
not been used here. We have only used the property of the stars in being
classified as solar analogues, and, consequently, we assumed their average
temperature to be equal to the solar effective temperature.

In our procedure, we are implicitly assuming that the correction in our
temperature scale is just a zero point offset and that no dependence on
temperature or metallicity is present.  These assumptions are justified
{\it a posteriori} in Sect. \ref{comparacions}, where several comparisons
of SEDF temperatures with other photometric and spectroscopic
determinations are shown.

\begin{figure}
\begin{center}
\leavevmode
\resizebox{\hsize}{!}{\includegraphics{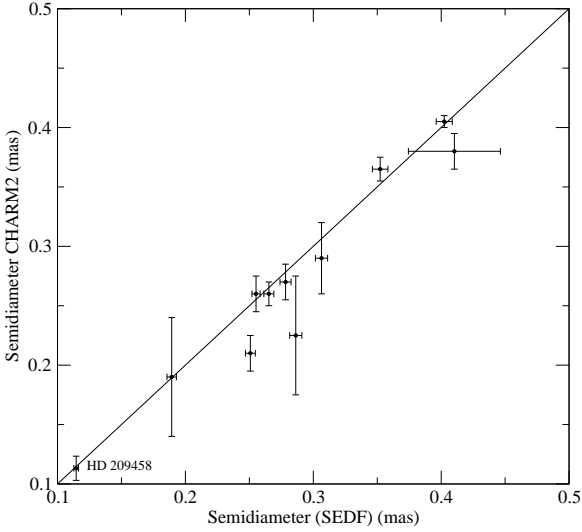}}
\end{center}
\caption{Comparison of angular semi-diameters computed from the SEDF
method (with the new zero point in the temperature scale) and from the
CHARM2 catalogue. In the case of HD 209458, the comparison of the
semi-diameter is between the SEDF method and an empirical determination
from a high-precision transit light curve.}
\label{figsemidia}
\end{figure}

The angular semi-diameters computed from Eq. (\ref {semidia}) were used to
check the consistency of the new zero points in our temperature
scale. These angular semi-diameters were compared with the direct values
compiled in the CHARM2 catalogue \citep{richichi05}. We restricted the
comparison to stars with accurate VLBI or indirect (spectrophotometry)
measurements of the semi-diameter. Only 10 of these stars fulfill the
conditions for applicability of the SEDF method. Figure \ref {figsemidia}
shows the comparison of the semi-diameters for these 10 stars. 
The agreement is excellent, with an average difference 
($\theta_{\rm dir}-\theta_{\rm SEDF}$), weighted
with the inverse of the square of the error, equal to 
$-$0.3\% with a s.d. of 4.6\%  (see
Table \ref{compsemi}). All the direct values used in the comparison
correspond to an uniform stellar disk. A crude comparison of both
uniform disk and limb darkened values for about 1600 F, G and K
stars in the CHARM2 catalogue indicates a $\sim$4\% positive correction
for limb darkening, of the same order of the dispersion of the relative 
differences shown in Table \ref{compsemi}. 
In addition, we compared the radius of HD 209458 --
obtained with the Hubble Space Telescope from a high precision planetary
transit light curve \citep{brown01} -- with our estimation from SEDF,
obtaining very good agreement: $1.146\pm0.050$~$R_{\odot}$ and
$1.160\pm0.058$~$R_{\odot}$, respectively.

\begin{table}
\caption {\sf Comparison of direct and SEDF angular semi-diameters.  
$\theta_{\rm dir}$ are VLBI and spectrophotometric values
\citep{richichi05}. For HD 209458, $\theta_{\rm dir}$ is derived from a
planetary transit light curve.}
\begin{center}
\begin{tabular}{lccr}
       \hline
 Star & $\theta_{\rm dir}$ (mas) & $\theta_{\rm SEDF}$ (mas) & $\Delta\theta (\%) $ \\
       \hline
       \hline
HIP   6702& 0.190$\pm$0.050& 0.189$\pm$0.004&    0.5\\
HIP   8433& 0.225$\pm$0.050& 0.286$\pm$0.005& $-$27.1\\
HIP  48113& 0.405$\pm$0.005& 0.402$\pm$0.006&    0.7\\
HIP  50786& 0.260$\pm$0.010& 0.265$\pm$0.004& $-$1.9\\
HIP  51056& 0.290$\pm$0.030& 0.306$\pm$0.005& $-$5.5\\
HIP  85365& 0.380$\pm$0.015& 0.410$\pm$0.036& $-$7.9\\
HIP  91237& 0.210$\pm$0.015& 0.251$\pm$0.004& $-$19.5\\
HIP  96895& 0.270$\pm$0.015& 0.278$\pm$0.004& $-$3.0\\
HIP  96901& 0.260$\pm$0.015& 0.255$\pm$0.003&    1.9\\
HIP 113357& 0.365$\pm$0.010& 0.352$\pm$0.006&    3.6\\
HD  209458& 0.113$\pm$0.010& 0.115$\pm$0.002& $-$1.8\\
\hline
\end{tabular}
\end{center}
\label{compsemi}
\end{table}

\subsection{Error estimation}
\label{errors}

One of the features of the SEDF method is that it yields individual
uncertainties of both \tefts and $\theta$. The total uncertainty can be
calculated by combining the contributions from the spectral energy
distribution fit and also from the uncertainties in the fixed parameters,
i.e., $\log g$, $[m/H]$, and $A_V$. Assuming null correlation between
these different (and, in principle, independent) sources of error, the
total uncertainty of the $Y_k$ parameter (\tefts or $\cal{A}$) is the
quadratic sum of the different error contributions:
\begin{eqnarray}
(\Delta Y_k)^2 &=& (\Delta {Y_k}^{\rm SEDF})^2+\left({\partial Y_k\over\partial 
[m/H]}\right)^2 (\sigma_{[m/H]})^2 \nonumber \\
& +& \left({\partial Y_k\over\partial \log g}\right)^2 
(\sigma_{\log g})^2 + \left({\partial Y_k\over\partial A_V}\right)^2 
(\sigma_{A_V})^2\nonumber\\
\label {eqError}
\end{eqnarray}
As mentioned above, the minimization of $\chi^2$ is carried out with
respect to the four magnitudes $VJHK$, with the parameters $A_V$, \feh and
\logg being held fixed. The first term of the equation ($\Delta {Y_k}^{\rm
SEDF}$) is the error in the parameter $Y_k$ coming from the fit to the
spectral energy distribution, which is computed from the covariance matrix
($\Delta {Y_k}^{\rm SEDF}\equiv\sqrt{C_{kk}}$). The derivatives in the
other three terms of the equation are determined numerically:
\begin{equation}
{\partial Y_k\over\partial [m/H]} \approx {{Y_k (+\Delta [m/H]) - Y_k 
(-\Delta [m/H])} \over {2}}
\end{equation}
\noindent in the case of the metallicity, and in an analogous way for
the surface gravity and the interstellar absorption.

The error in \tefts is obtained directly from Eq. (\ref{eqError}), whereas
the error in $\theta$ must be calculated from the error in the $\cal{A}$
parameter:
\begin{equation}
\sigma_{\theta} = 0.2 \,\ln 10\,\theta\,\sigma_{\cal{A}}
\label {errTheta}
\end{equation}
The adopted values for the errors in the magnitudes ($\sigma_{m_i}$),
metallicity ($\sigma_{[M/H]}$) and surface gravity ($\sigma_{\log g}$) are
discussed in Sect. \ref{errSample}.

\section{Comparison with other methods}
\label{comparacions}

Five samples of FGK stars with accurate determinations of effective
temperatures (both photometric and spectroscopic) were selected from the
literature (\citealp{alonso96a}, \citealp{ramirez05a}, \citealp
{fuhrmann98}, \citealp {santos03} and \citealp{edvardsson93}) to carry out
a comparison with our results. We put special attention in correcting for
the effects of interstellar reddening, which could lead to systematic
differences. For the \citeauthor{alonso96a} and \citeauthor{ramirez05a}
samples (the most reddened), interstellar reddening was corrected using
the values of $E(B-V)$ given by the authors so that the two temperature
estimations would be directly comparable. For the other three samples,
composed of stars at closer distances, we restricted our comparisons to
unreddened stars. Anyhow, this meant rejecting very few stars from further
analysis. Among several papers in the literature, we chose
these five samples because they have a minimum of 25
stars with non-saturated \MASS photometry and the
values of \feh and \logg -- needed for a consistent comparison -- are
provided by the authors.

\subsection{Methods based on IR photometry: \citet{alonso96a} and
\citet{ramirez05a}}

\begin{figure}[t]
\resizebox{\hsize}{!}{\includegraphics{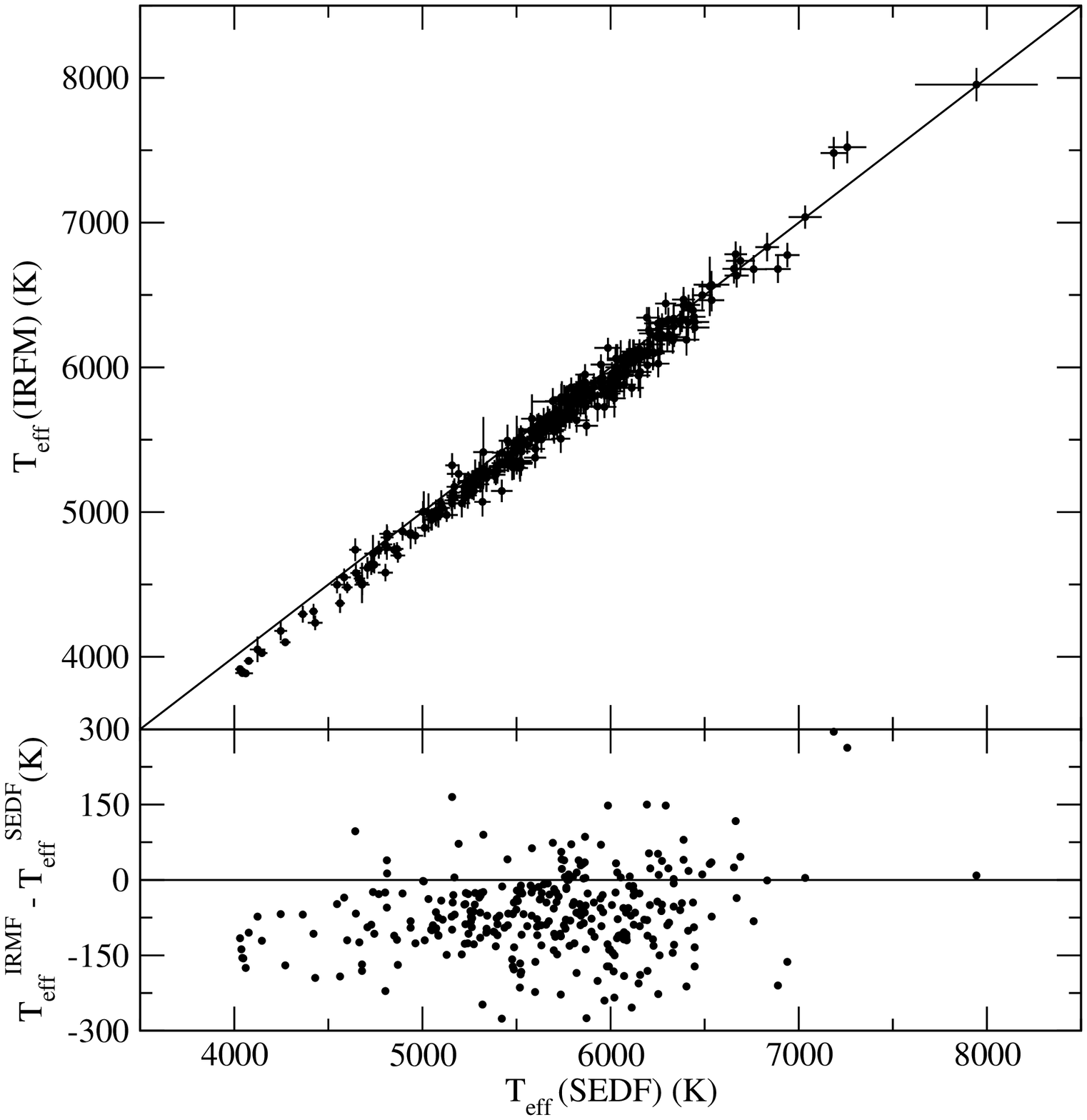}}
\resizebox{\hsize}{!}{\includegraphics{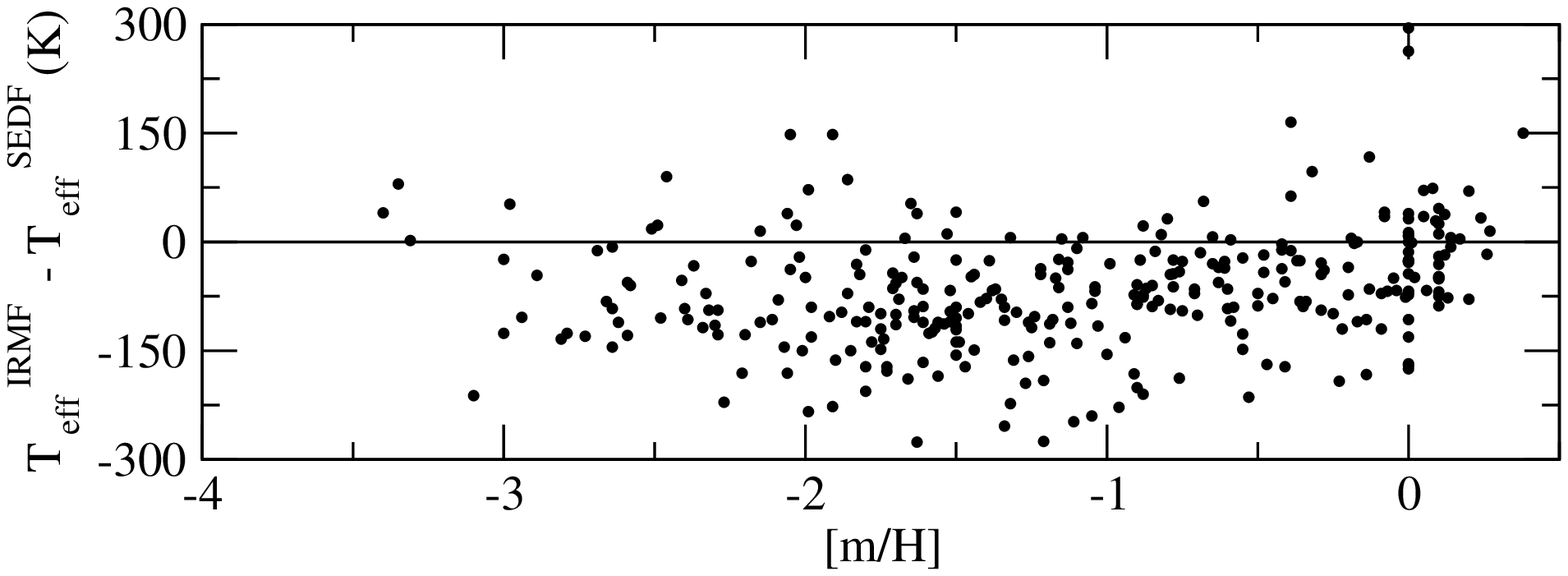}}
\caption{Comparison of the effective temperatures from the IRFM and the
SEDF method for 315 stars in the sample of \citet {alonso96a}. The bottom
panel shows the temperature difference as a function of the metallicity.}
\label{figAlonsoT}
\end{figure}

As mentioned above, the IRFM is the most popular method to compute
effective temperatures from IR photometry. The work by \citet{alonso96a}
is undoubtedly the widest application of the IRFM to FGK stars. The
authors computed effective temperatures for 462 stars with known
interstellar absorption, surface gravity and metallicity. After selecting
the stars in the \citeauthor{alonso96a} sample in the range $4000 < \tef <
8000$ K and with errors in the \MASS photometry below 0.05 mag, we
obtained effective temperatures from the SEDF method for a subset of 315
stars. The comparison between both determinations of \tefts is shown in
Fig.~\ref{figAlonsoT}. The average difference $\Delta \tef$ (IRFM $-$
SEDF) was found to be $-$67~K, with a standard deviation of 81~K. The
dependence of this difference on the temperature is not significant:
$T_{\rm eff}^{\rm IRFM} = 1.030\: T_{\rm eff}^{\rm SEDF} - 239$ K. As
shown in the bottom panel of Fig.~\ref{figAlonsoT}, there is no dependence
of the temperature difference with metallicity.

\begin{figure}[t]
\resizebox{\hsize}{!}{\includegraphics{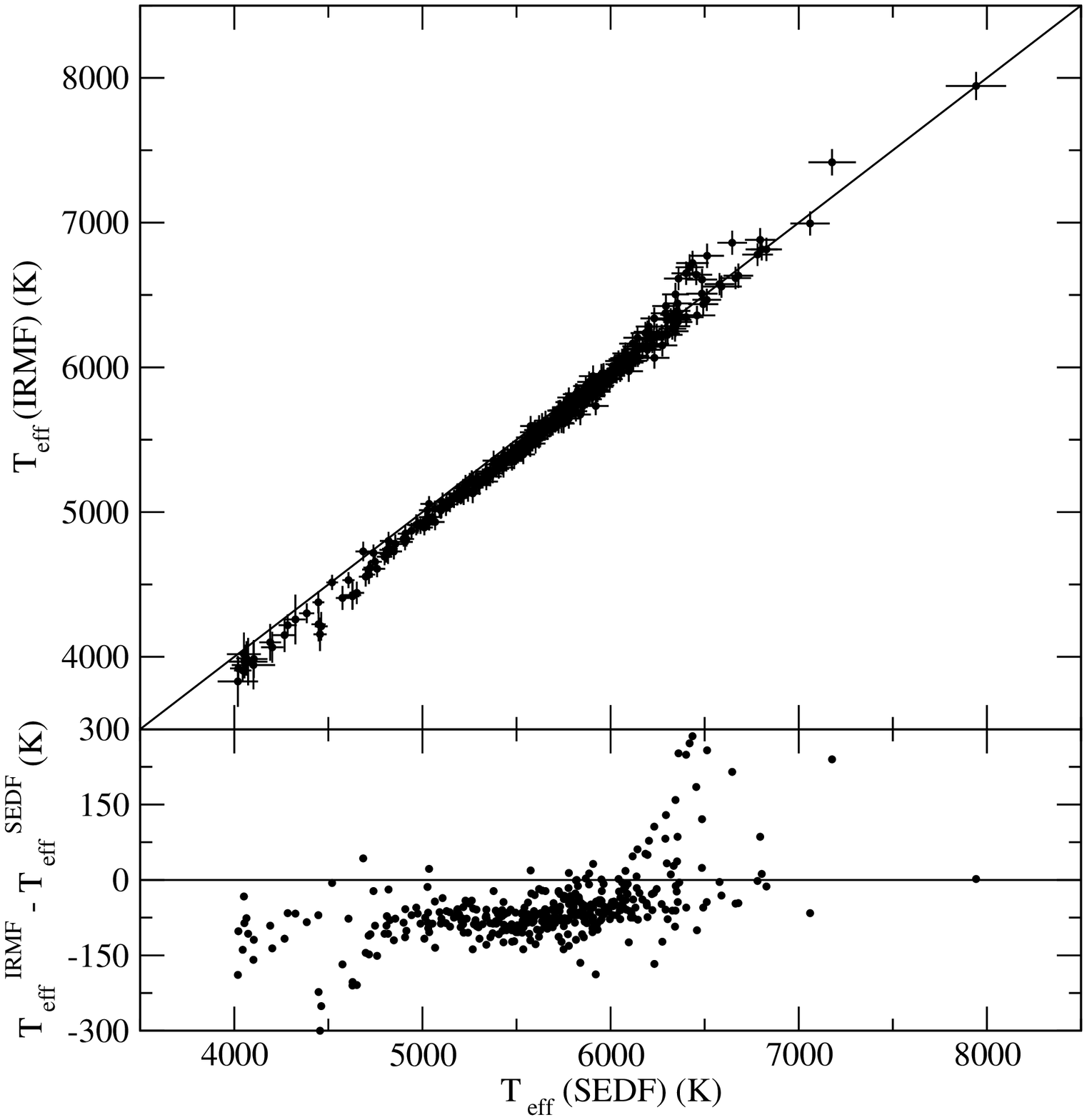}}
\resizebox{\hsize}{!}{\includegraphics{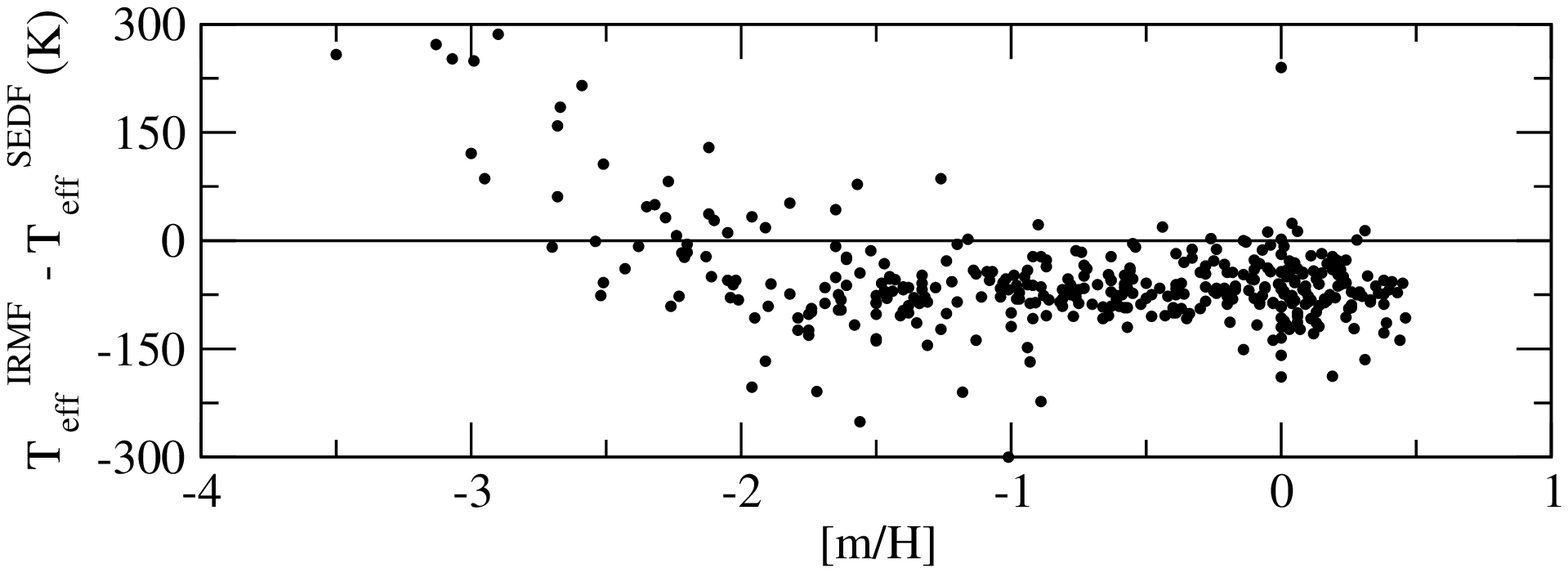}}
\caption{Comparison of the effective temperatures from the IRFM and the
SEDF method for 386 stars in the sample of \citet {ramirez05a}. The bottom
panel shows the temperature difference as a function of the metallicity.}
\label{figRamirezT}
\end{figure}

In a recent work, \citet{ramirez05a} have recomputed the IRFM
temperatures of almost all the stars in \citet{alonso96a} using updated
input data. According to the authors, the difference between the old and
new temperature scales is not significant. They also compare their
effective temperatures with some direct determinations. 
The authors conclude that there is a systematic difference of about 40 K
at solar temperature (in the sense IRFM$_{\rm Alonso}$ -- their values).  
The comparison between the temperatures of \citeauthor{ramirez05a} and our
determinations is shown in Fig.~\ref{figRamirezT}. For 385 stars in
common we find $\Delta \tef$ (IRFM $-$SEDF) equal to $-$58~K 
($\sigma_{T_{\rm eff}}$=67 K), and $T_{\rm
eff}^{\rm IRFM} = 1.061\: T_{\rm eff}^{\rm SEDF} - 403$ K. Unlike in the
case of \citet{alonso96a}, the dependence of $\Delta \tef$ in \feh is
relevant (Fig.~\ref{figRamirezT}, bottom panel). For \feh $<-2.0$ the
temperatures from \citeauthor{ramirez05a} are clearly hotter than our
temperatures. The same trend was found by \citet{charbonnel05} when
comparing their temperatures of 32 halo dwarfs ($-3.5 < [Fe/H] < -1.0$)
with the values of \citeauthor{ramirez05a}.
 
\subsection{Other methods}

Besides the IRFM, which uses IR photometry, we have also compared the
effective temperatures obtained using the SEDF method with other
determinations. Two of these (\citealp{fuhrmann98} and \citealp{santos03})
are spectroscopic works, while in another case \citep{edvardsson93} the
temperatures are based on \uvbys photometry. The chief problem in the case
of spectroscopic determinations is that, in general, they are mostly
applied to bright stars, which have poor \MASS photometry (the \MASS
detectors saturate for stars brighter than $K\approx 4$ mag). This fact
reduces the number of stars in the \citet{fuhrmann98} and \citet{santos03}
samples that can be compared with SEDF method.

\subsubsection{\citet{fuhrmann98}}

\begin{figure}[t]
\resizebox{\hsize}{!}{\includegraphics{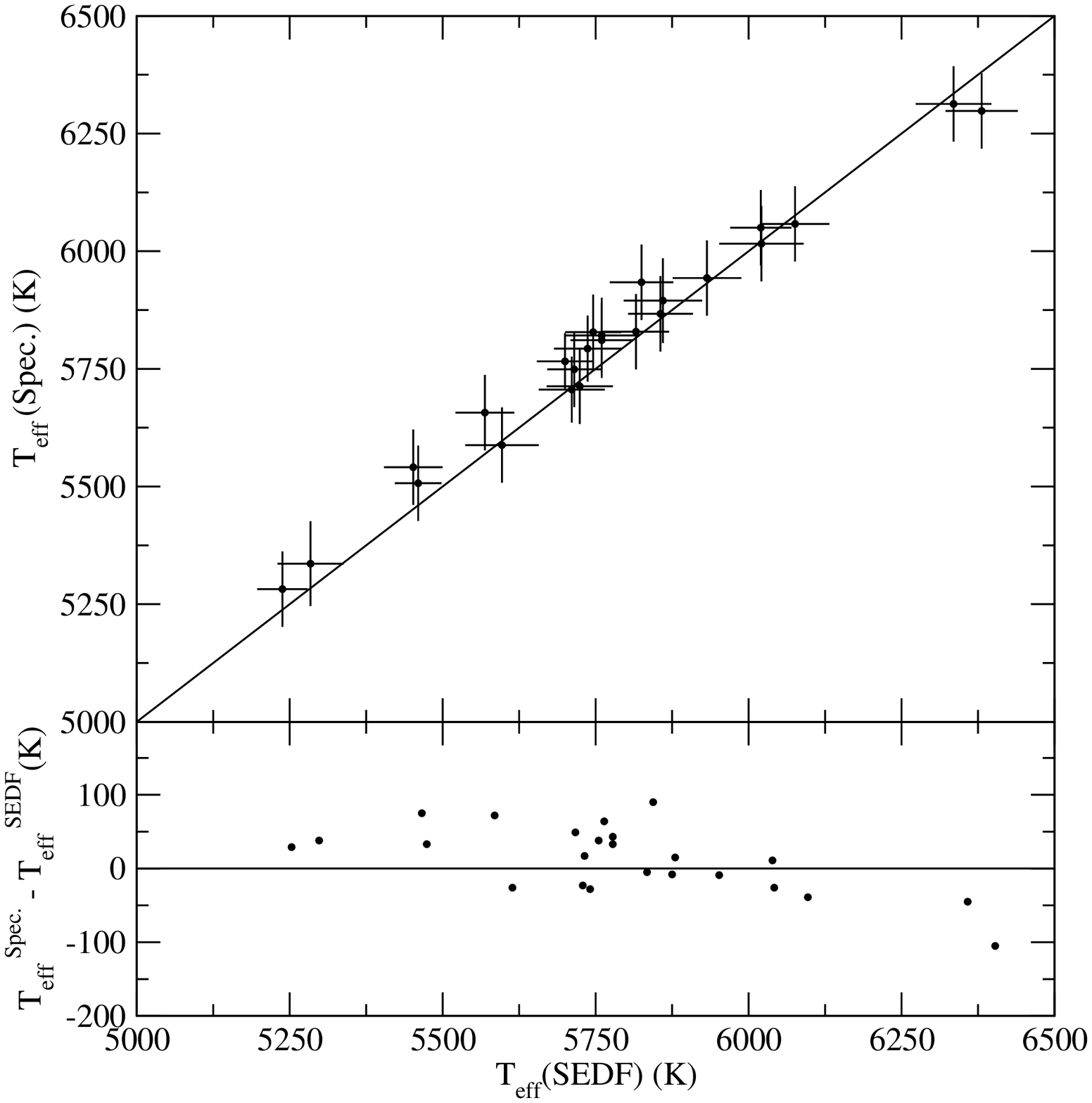}}
\resizebox{\hsize}{!}{\includegraphics{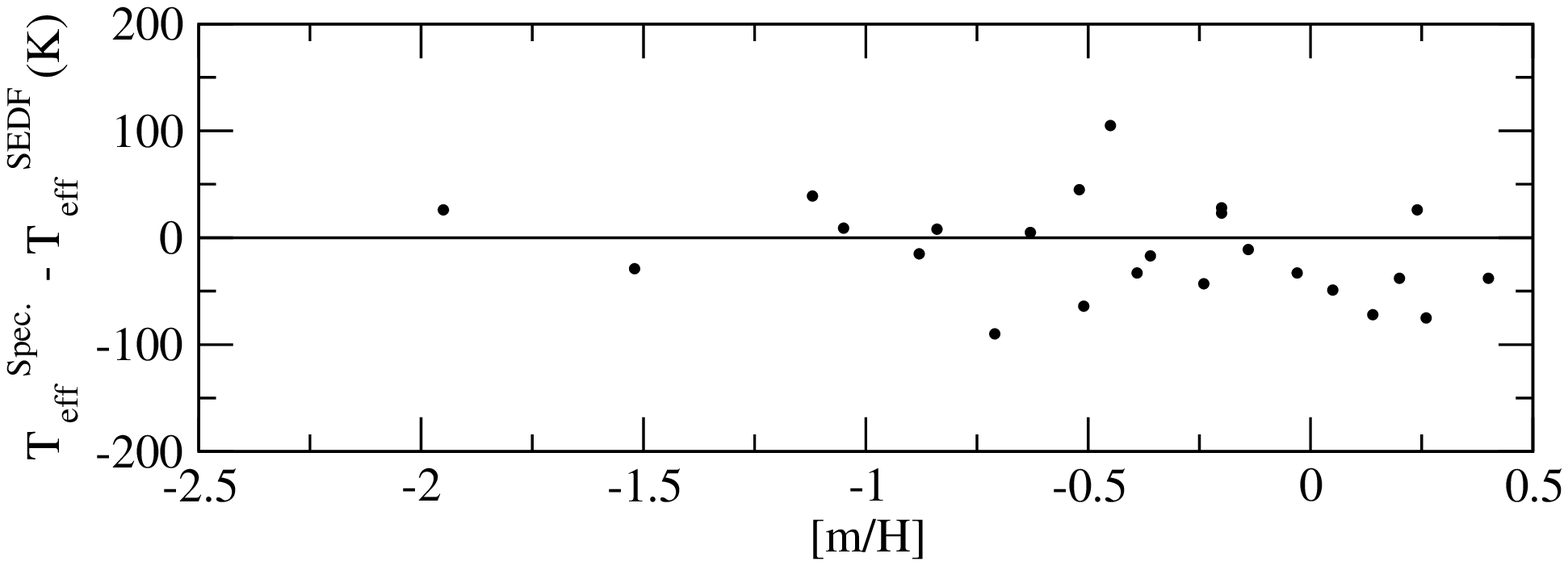}}
\caption{Comparison of the effective temperatures from fits to Balmer
lines and the SEDF method for 24 stars in common with the sample of \citet
{fuhrmann98}. The bottom panel shows the temperature difference as a
function of the metallicity.}
\label{figFuh}
\end{figure}

This sample is composed of about 50 F and G nearby stars, both main
sequence and subgiants, of the Galactic disk and halo.  Effective
temperatures were determined from fits to the wings of the Balmer lines.
Of those stars, 24 have accurate \MASS photometry so that reliable SEDF
temperatures can be derived. The comparison is shown in Fig. \ref{figFuh}.
The mean average difference $\Delta T_{\rm eff}$ (Fuhrmann $-$ SEDF) is 12
K, ($\sigma_{T_{\rm eff}}$=45 K), with a slight dependence on the
temperature: $T_{\rm eff}^{\rm Fuhrmann}=0.895\: T_{\rm eff}^{\rm SEDF} +
618$ K. No dependence was found between $\Delta T_{\rm eff}$ and \feh
(Fig.~\ref{figFuh}, bottom panel).

\subsubsection{\citet{santos03}}

\begin{figure}[t]
\resizebox{\hsize}{!}{\includegraphics{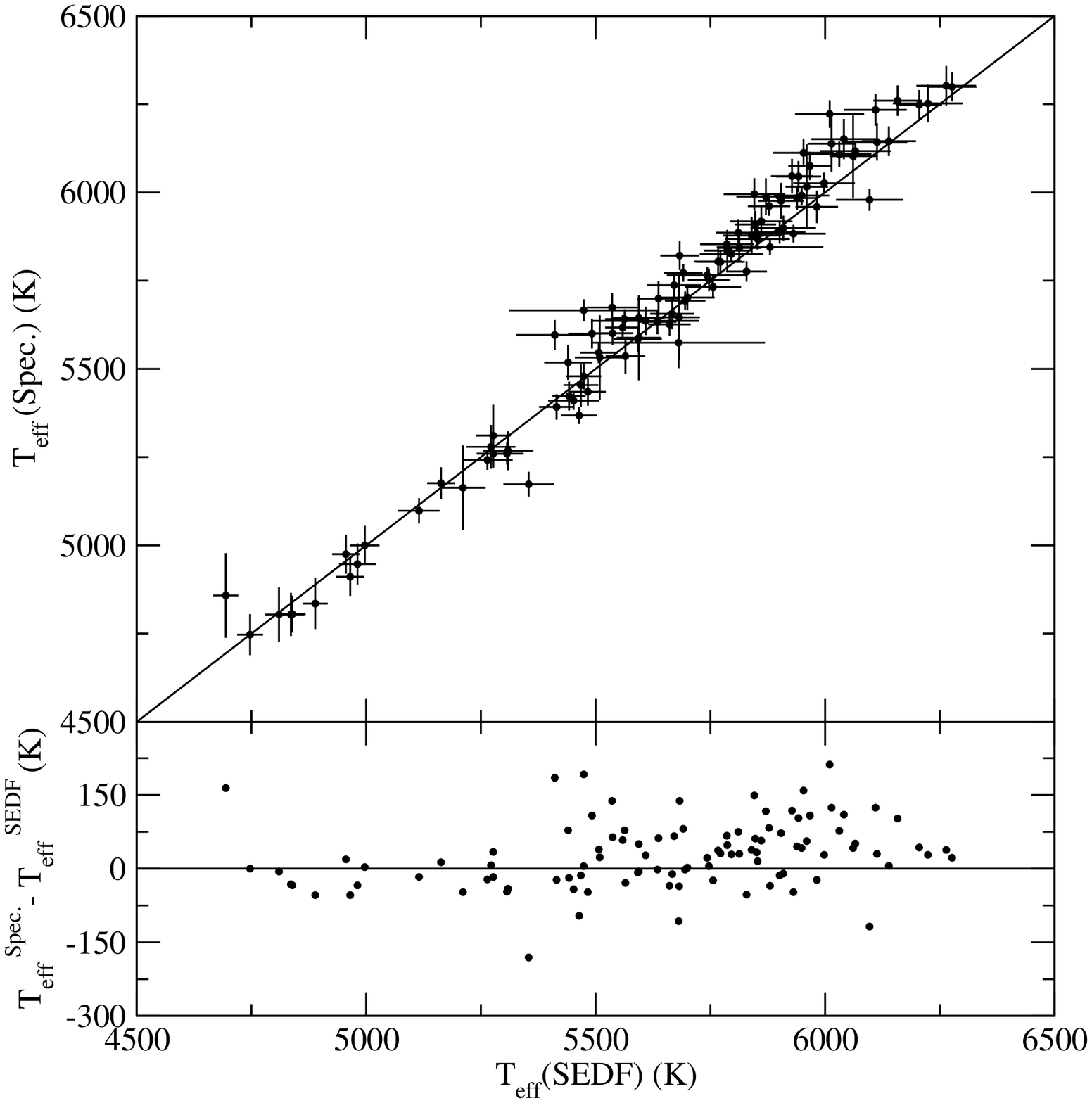}}
\resizebox{\hsize}{!}{\includegraphics{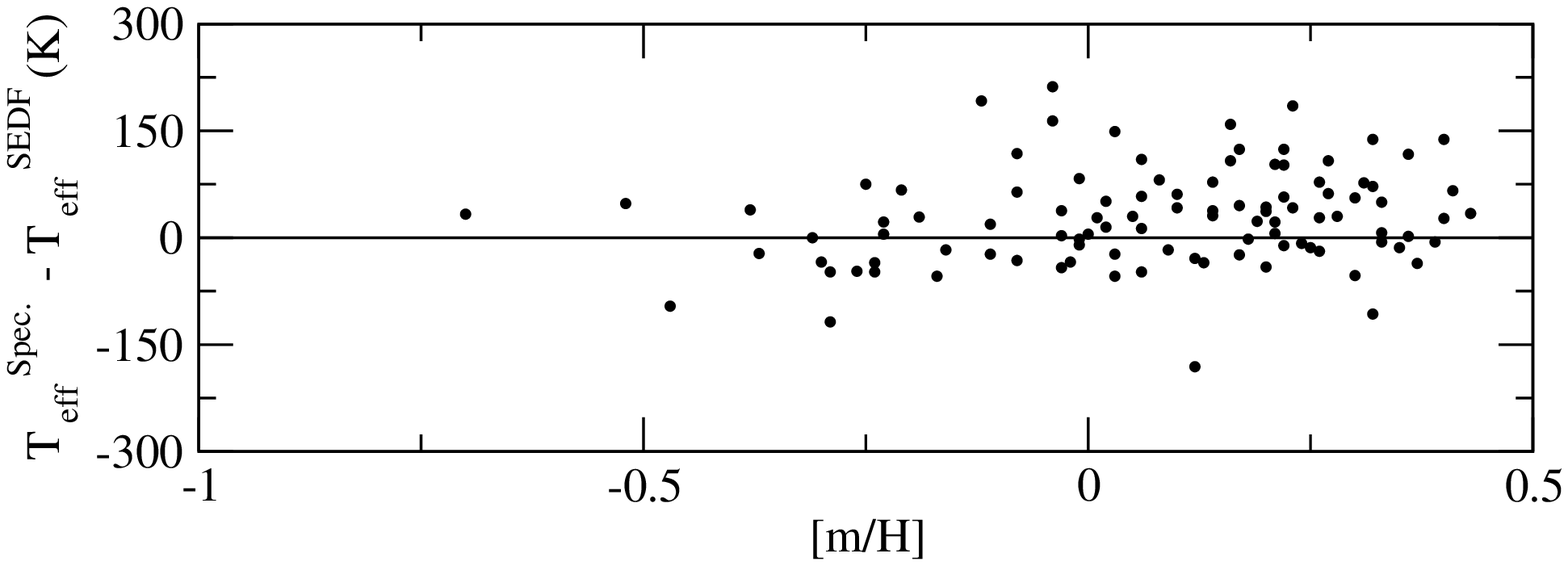}}
\caption{Comparison of the effective temperatures from iron line fits and
the SEDF method for 101 stars in common with the sample of \citet
{santos03}. The bottom panel shows the temperature difference as a
function of the metallicity.}
\label{figSan}
\end{figure}

To study the correlation between the metallicity and the probability of a
star to host a planet, \citet{santos03} obtained spectroscopic
temperatures for 139 stars based on the analysis of several iron lines.
Effective temperatures for a total of 101 stars in
the sample of \citeauthor{santos03} can be obtained using the SEDF method.
In this case, $\Delta T_{\rm eff}$ (Santos $-$ SEDF) is 28~K, with
$\sigma_{T_{\rm eff}}=68$ K, and practically independent of the
temperature: $T_{\rm eff}^{\rm Santos} = 1.053\: T_{\rm eff}^{\rm SEDF}
-270$ K (Fig. \ref{figSan}). There is no dependence of $\Delta T_{\rm
eff}$ with \feh (Fig.~\ref{figSan}, bottom panel).

\subsubsection{\citet{edvardsson93}}

\begin{figure}[t]
\resizebox{\hsize}{!}{\includegraphics{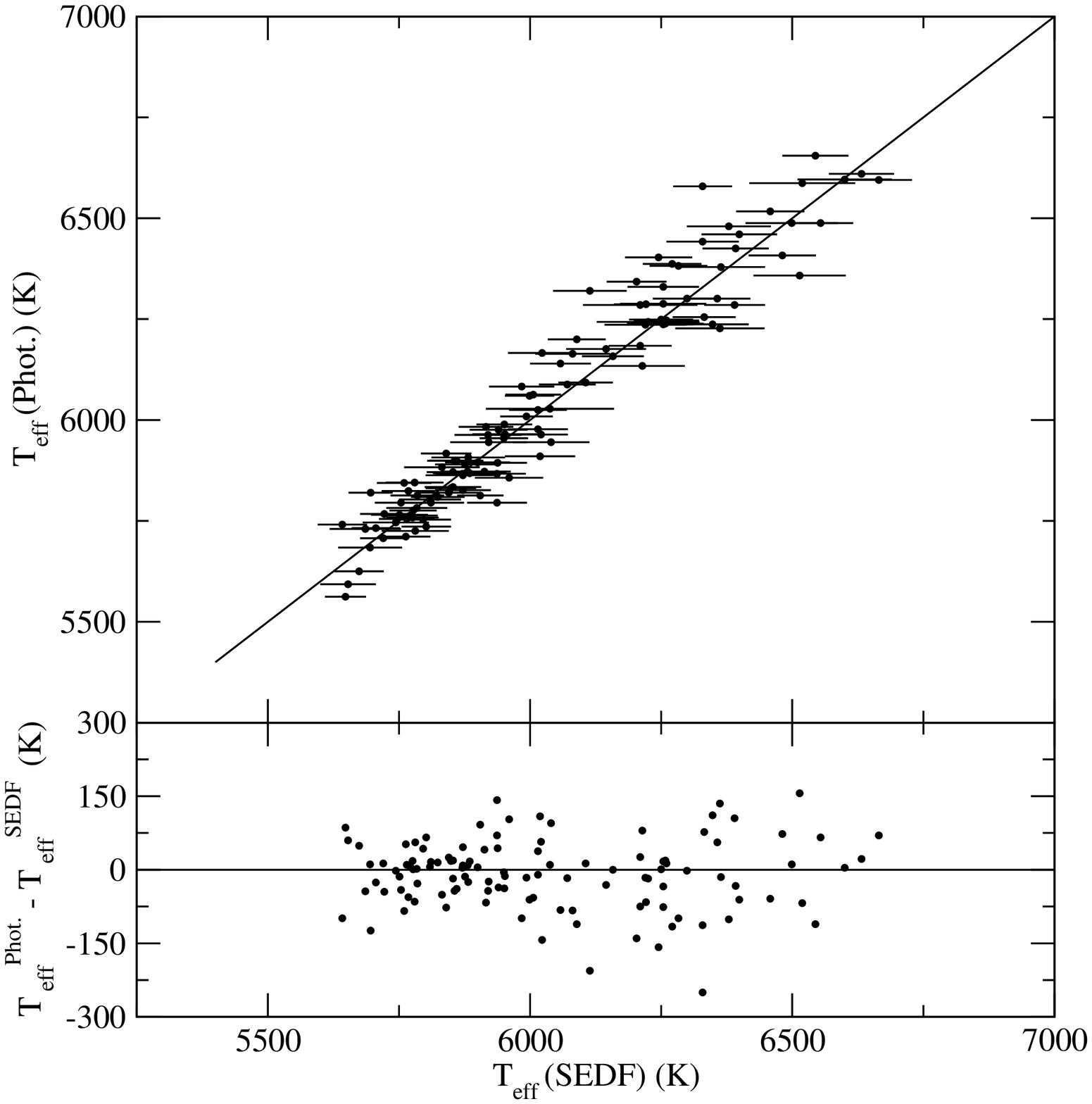}}
\resizebox{\hsize}{!}{\includegraphics{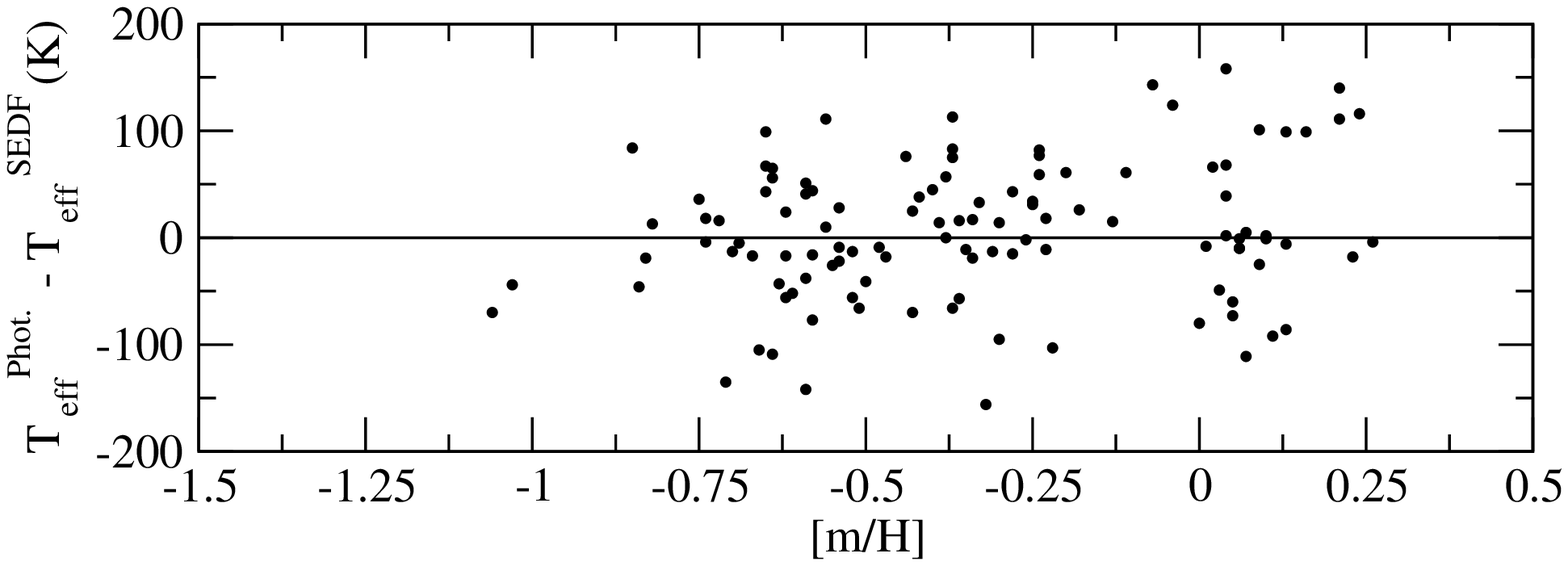}}
\caption{Comparison of the effective temperatures from photometry and the
SEDF method for 115 stars in common with the sample of
\citet{edvardsson93}. The bottom panel shows the temperature difference as
a function of the metallicity. The standard deviation for a single star
in \citet{edvardsson93} is 81~K.}
\label{figEdv}
\end{figure}

The sample of \citeauthor{edvardsson93} is composed by 189 nearby F and G
type stars. In contrast with the previous two, the effective temperature
is not derived from spectroscopy but from \uvbys photometry. To do so, the
authors built a grid of synthetic photometry using the atmosphere models
of \citet{gustafsson75} and further improved it by adding several new
atomic and molecular lines. Effective temperatures for 115 stars in their
sample could be derived using the SEDF method. The average difference
$\Delta T_{\rm eff}$ (Edvardsson $-$ SEDF) is 10 K, with a dispersion of
70 K and no dependence on the temperature: $T_{\rm eff}^{\rm Edvardsson} =
1.006\: T_{\rm eff}^{\rm SEDF} -27$ K (Fig. \ref{figEdv}). As in the two
previous cases, the bottom panel of Fig.~\ref{figEdv} shows that the
temperature difference is not correlated with \FEH.

\section{Parametric calibrations}
\label{calib}

The practical use of the SEDF method as it has been described in Sect.
\ref{sedf} is not straightforward since it requires the calculation of
synthetic photometry from stellar atmosphere models and then use a
numerical algorithm to minimize the $\chi^2$ function. Parametric
calibrations (as a function of one or more parameters) may offer a
suitable means to estimate reliable effective temperatures in cases where
simplicity and speed are to be preferred over the best possible accuracy.
In this section we present calibrations for both \tefts and \bc as a
function of $(V-K)_0$, \FEH and \logG.  To calculate the
calibrations, the SEDF method was applied to a sample of stars in the
Hipparcos catalogue, as described below. Note that these calibrations are
subject to two limitations with respect to the full SEDF method: First,
they are simplifications since not all the available information is used,
and second, individual uncertainties cannot be determined.

\subsection{The stellar sample}
\label{sample}

We collected a sample of FGK dwarfs and subdwarfs in the Hipparcos
catalogue, and therefore with measured trigonometric parallaxes.  Their
$V$ magnitudes come mainly from the \citet{hauck98} catalogue, except for
those stars with less than two observations, where we used the Hipparcos
catalogue. The entire sample has complete and non-saturated $JHK$
photometry in the \MASS catalogue. The metallicity was extracted from the
compilation of \citet{cayrel01} or computed from $uvby-\beta$ photometry
-- either measured from our own observations or obtained from the
\citet{hauck98} catalogue --, using a slightly revised version of the
\citet{schuster89b} calibration. The range of metallicities covered by the
sample is $-3.0 \lesssim [m/H] \lesssim 0.5$. Values of \logg were
computed from $uvby-\beta$ photometry \citep{masana94, jordi96}.  
Originally, the sample was built to study the structure and kinematics of
the disk and halo of the Galaxy (\citet{masana04}) and a full description
including the photometry and a complete set of physical parameters will be
provided in a forthcoming paper \citep{masana06}.

In spite of the proximity of the stars (90\% of them are closer than 200
pc), we computed individual interstellar absorptions from $uvby-\beta$
photometry and corrected the observed magnitudes. As discussed below,
interstellar absorption is one of the most important sources of
uncertainty in the \tefts determination.

\subsubsection{Errors}
\label{errSample}

\begin{figure*}
\begin{center}
\leavevmode
\includegraphics[width=6.0cm]{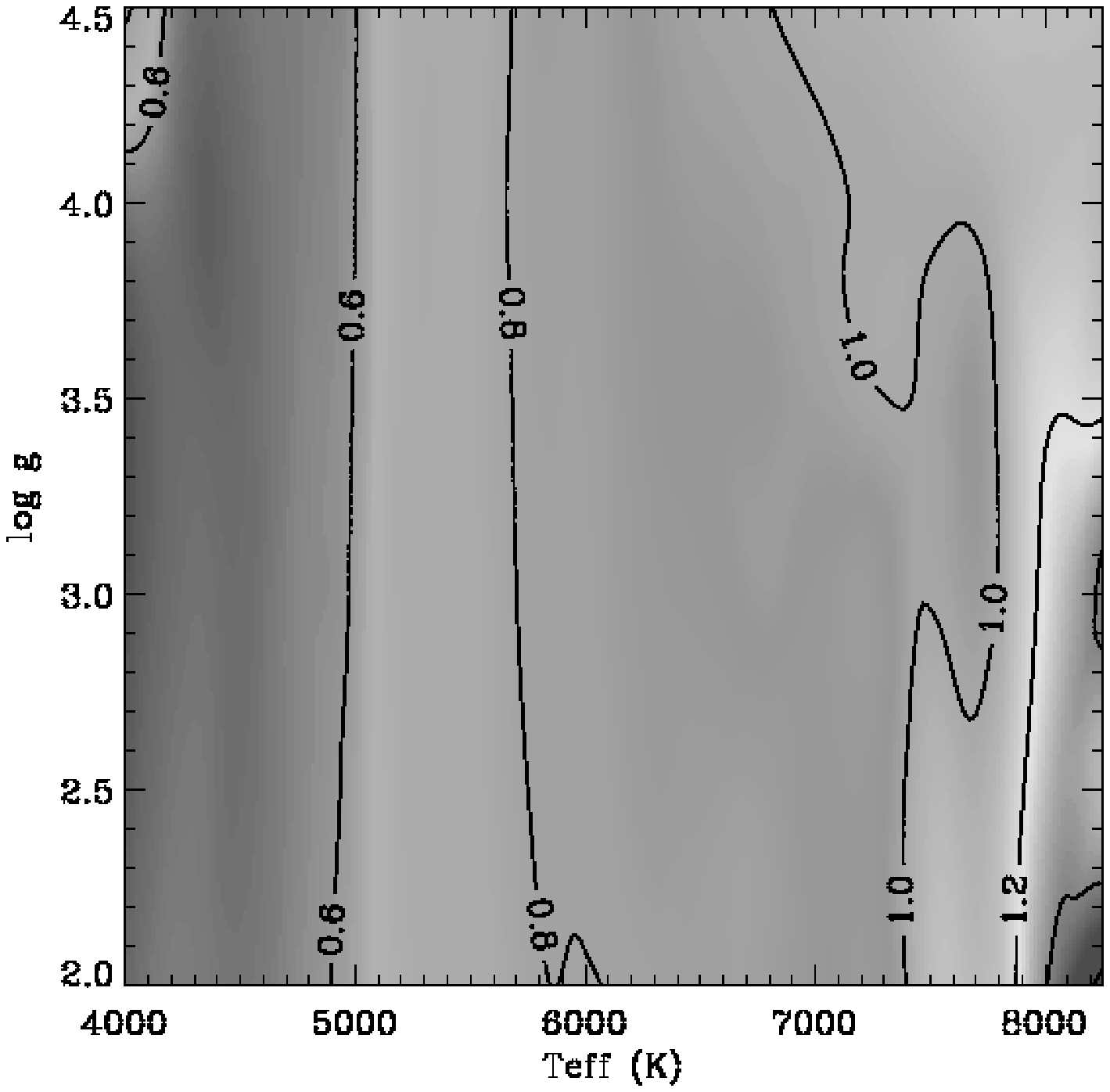}
\includegraphics[width=6.0cm]{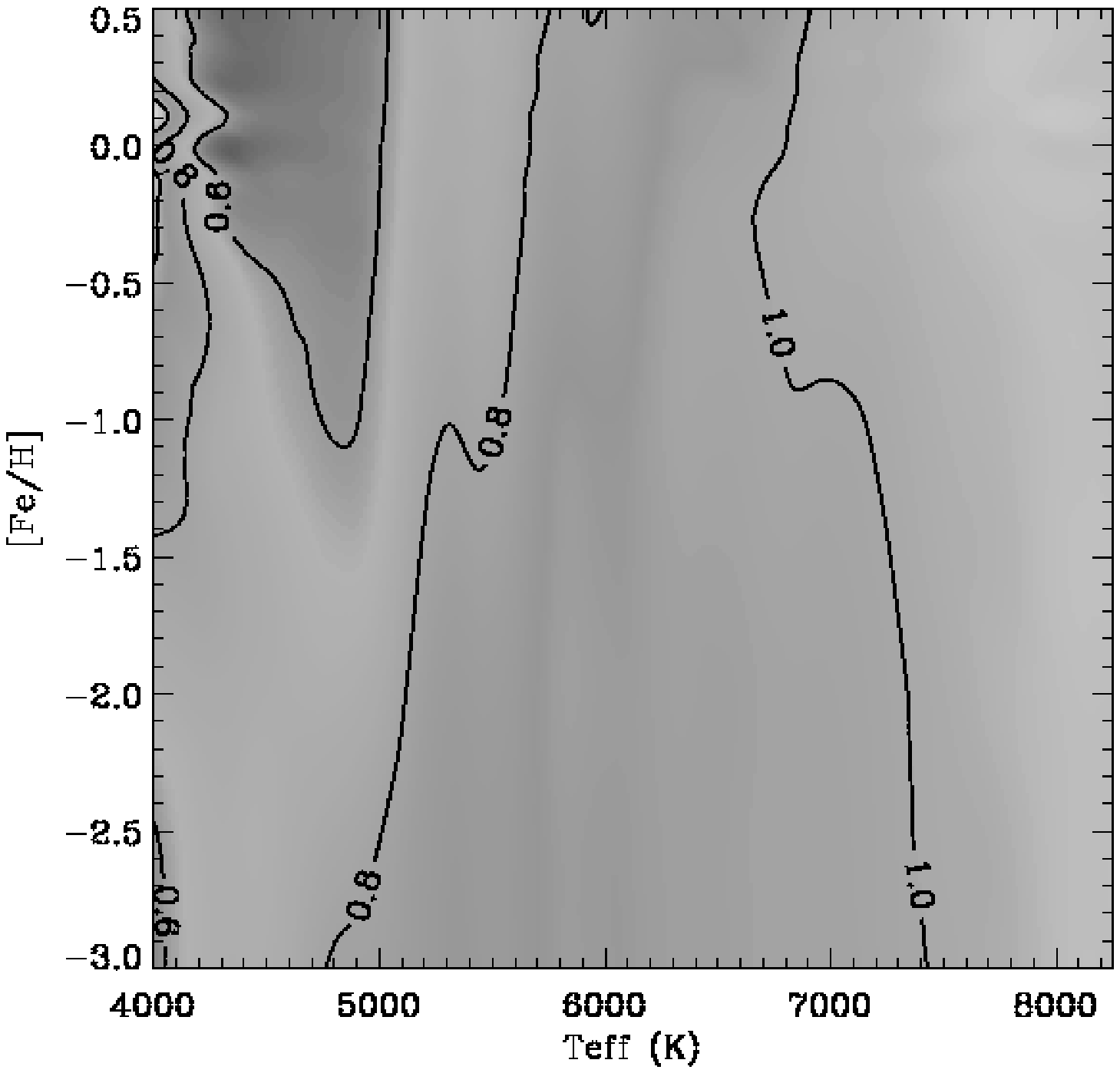}
\includegraphics[width=6.0cm]{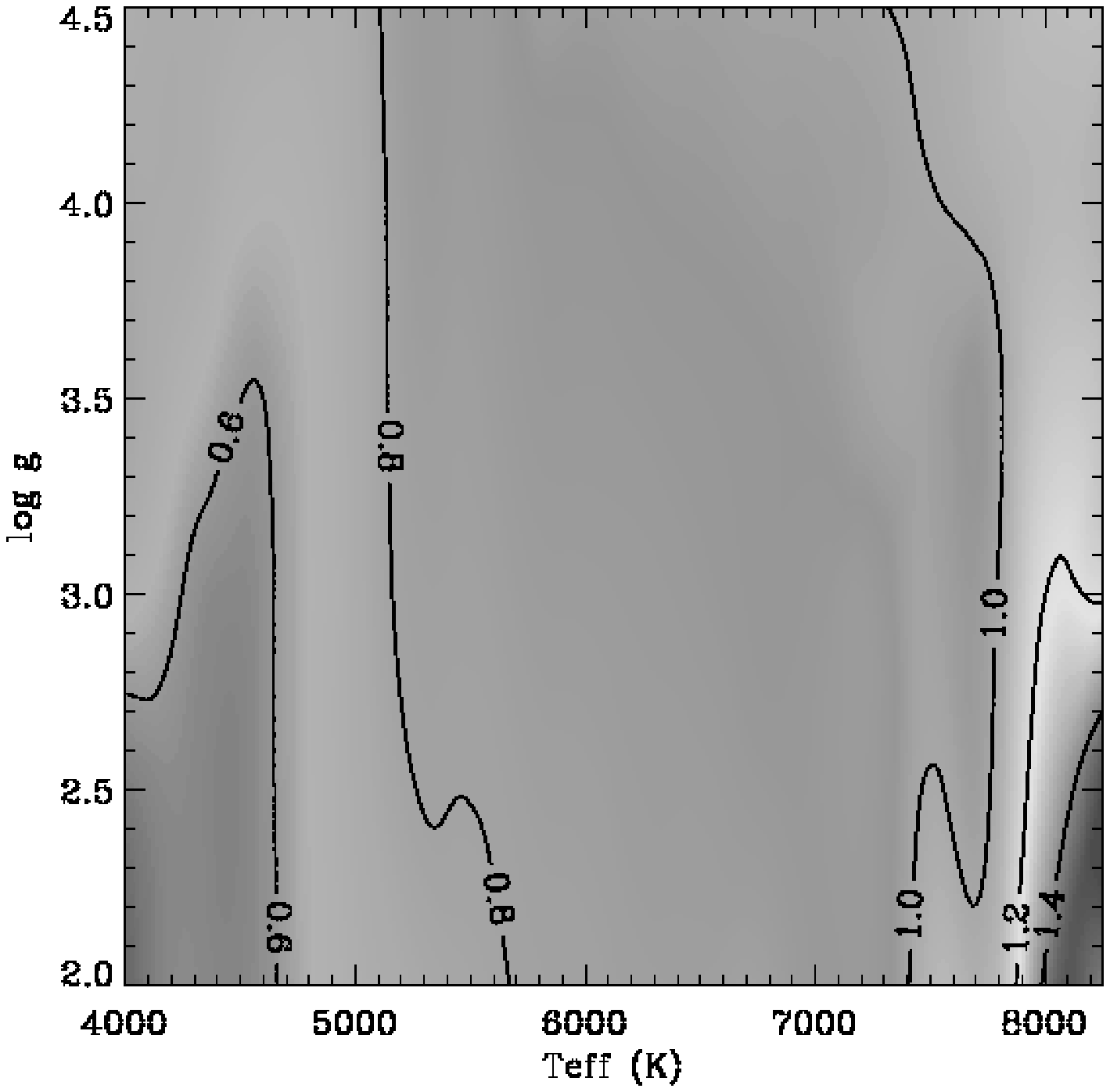}
\includegraphics[width=6.0cm]{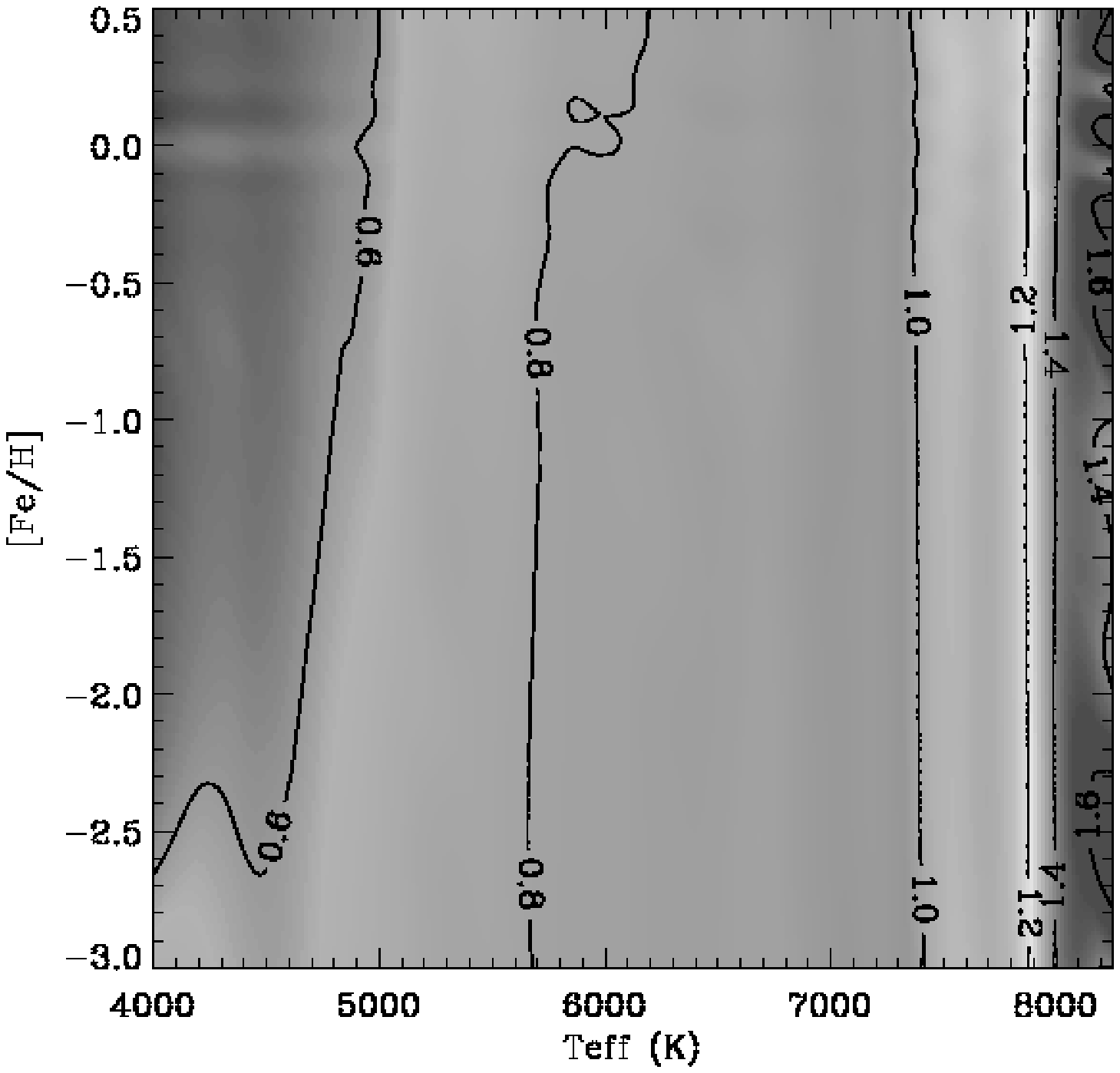}
\end{center}
\caption{Relative error (\%) in effective temperature assuming 
$\sigma_{VJHK}=0.015$ mag, $\sigma_{[m/H]}=0.2$, $\sigma_{\log g}=0.2$ 
and the error of the absolute flux calibration. 
Top left: $[m/H]=0.0$. 
Top right: $\log g=4.5$.
Bottom left: $[m/H]=-2.0$. 
Bottom right: $\log g=2.0$.}
\label{figErrTot}
\end{figure*}

For our sample, the errors in the magnitudes, metallicity and surface 
gravity were estimated in the following manner:
\begin{itemize}
\item {\em Errors in the $VJHK$ magnitudes:} The total error in each
magnitude was computed as the quadratic sum of the observational error,
the error in the absolute flux calibration and the error in the
determination of the interstellar extinction. The first one comes from the
photometric catalogues. However, to prevent the underestimation of the
error in the $V$ band, usually computed from the average of a few
measurements (and with no evaluation of systematics), we have set a
minimum error in $V$ equal to 0.015 mag. The uncertainties in the absolute
flux calibration are given by \citet{cohen03a,cohen03b} and are in the
range 1.5--1.7\% (0.016--0.019 mag), depending on the band. For those
stars affected by interstellar reddening, the uncertainty in $A_V$ as
derived from photometric calibrations based on \uvbys photometry
\citep{jordi96} is expected to be of about 0.05 mag, or $\sim$1.5\% in
\teft.
\item {\em Errors in \feh and \logG:} As mentioned above, \feh was
obtained, whenever possible, from spectroscopic measurements, and
otherwise we used photometric calibrations, with assigned uncertainties of
0.10 dex and 0.15 dex, respectively.  We assigned uncertainties of 0.18
dex to \logg values determined from photometric calibrations. The effect
on the final effective temperatures due to the uncertainties of both \feh
and \logg is very small: an error of 0.5 dex in \feh has an effect in
\tefts of less than 0.5\%, whereas the same error in \logg has an effect
that ranges between 0\% and 1\%, depending on the value of \tefts and
\logG.
\end{itemize}
No error was attributed to the flux in the stellar atmosphere models.
Comparisons carried out by using other stellar atmosphere models such as 
those by \citet{castelli97} and the NextGen models by \citet{hauschildt99} 
show resulting differences in temperature below $\sim 0.3\%$ in all cases
\citep{iribas03a}.

An estimation of the final errors in \tefts as function of \teft, \feh and
\logg is shown in Fig.~\ref{figErrTot}. As can be seen, the final error is
almost independent of \feh and \logG, but not of \teft. Hotter stars have
greater uncertainties (slightly $>$1\% for $\tef = 7500$ K) than cooler
stars (0.6\% for $\tef = 5000$ K). It is important to note that, in the
case of reddened stars, an uncertainty of 0.05 mag in $A_V$ can double the
error in \tefts compared to the values in Fig.~\ref{figErrTot}. For the
angular semi-diameter the behaviour of the errors is very similar to those
of the effective temperature, with values for unreddened stars of about
1.0--2.5\%. This means that for Hipparcos stars with good parallaxes, we
are able to determine the stellar radii with remarkable uncertainties of
about 1.5--5.0\%.

\begin{figure}[t]
\resizebox{\hsize}{!}{\includegraphics{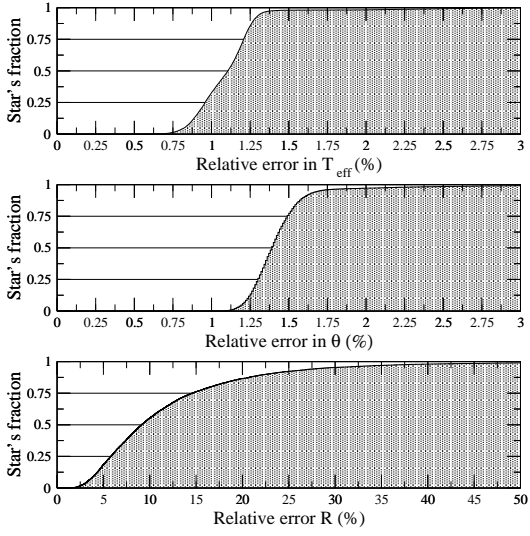}}
\caption{Cumulative histograms of the relative error in effective
temperature, angular semi-diameter and radius for the 10999 stars in the
sample.}
\label{figErrSample}
\end{figure}

Figure \ref{figErrSample} shows the cumulative histograms of the relative
errors in effectiVe temperature, angular semi-diameter and radius for the
10999 stars of the sample. As can be seen, about 85\% of the stars have
determinations of \tefts better than 1.1\%. The relative error in the
angular semi-diameter is also better than 1.5\% for about 85\% of the
stars. In the case of the radii, the main contributor to the error is the
uncertainty in the parallax. Even so, 50\% of the stars have radius
determinations better than 10\%, and 85\% of the stars better than 25\%.
It should be kept in mind that most of the stars in our
sample are essentially unreddened, thus yielding the best possible
accuracy.

Table \ref{tabsample} lists effective temperatures, angular
semi-diameters, radii and bolometric corrections in the $V$ and $K$ (2MASS)
bands with the corresponding uncertainties for the entire sample.
Using these values, we calculated simple parametric calibrations of
effective temperature and bolometric correction as described below.

\subsection{Effective temperature calibration}
\label{tempCal}

\begin{figure}
\begin{center}
\resizebox{\hsize}{!}{\includegraphics{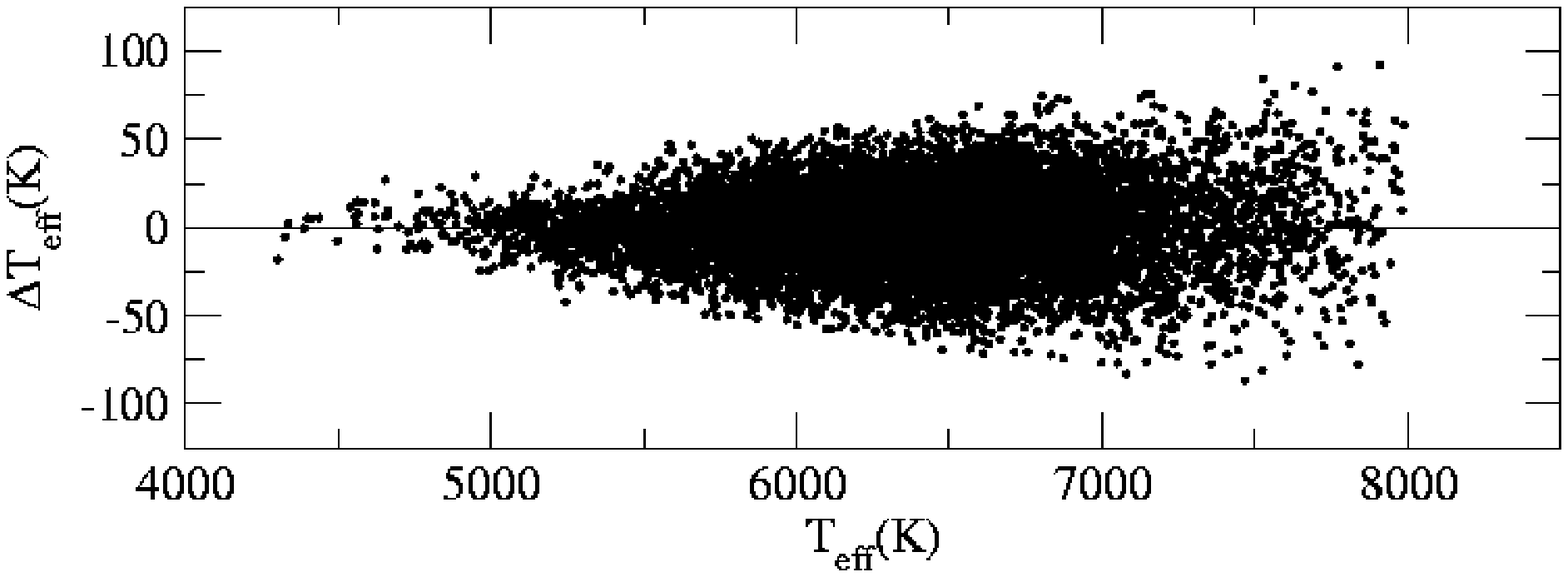}}
\resizebox{\hsize}{!}{\includegraphics{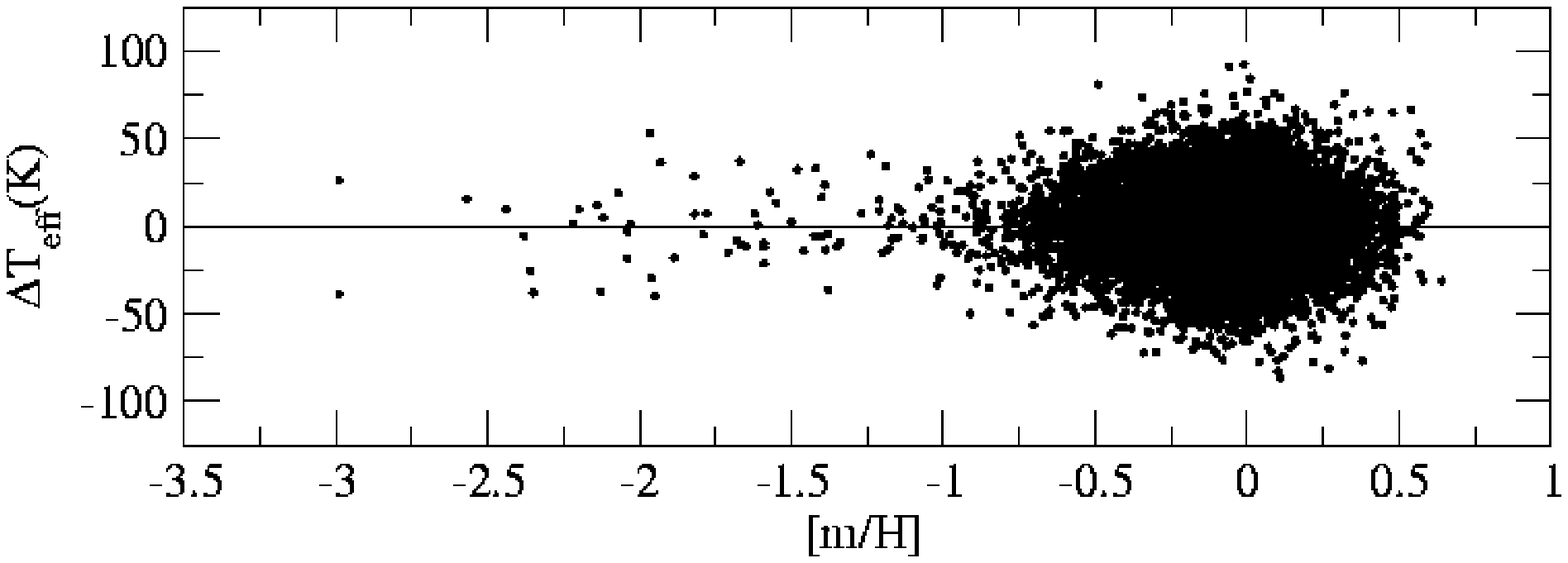}}
\resizebox{\hsize}{!}{\includegraphics{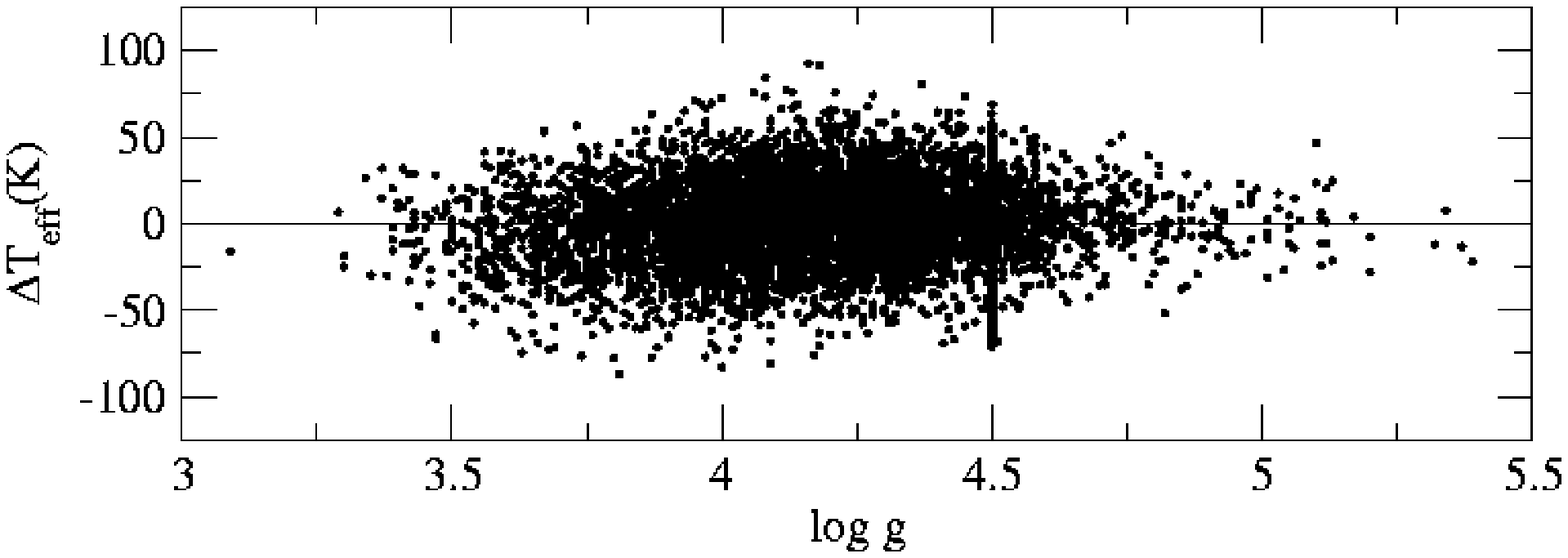}}
\end{center}
\caption{Residuals of the $T_{\rm eff}$ fit as function of effective 
temperature, metallicity and surface gravity.}
\label{resTempVK}
\end{figure}

Although the effective temperature for FGK type stars is strongly
correlated with the $(V-K)_0$ index (see for instance \citet{alonso96a}),
it also depends weakly on the metallicity and surface gravity, as we
mentioned in Sect. \ref{sedf}. Therefore, an empirical calibration of
\tefts should include terms in all $(V-K)_0$, \feh and \logG. Furthermore,
in our case the calibrations were constructed separately in two $(V-K)_0$
intervals. Stars departing more than $3\sigma$ from the fit were rejected.
The resulting expressions are:

\begin{itemize}
\item \hspace{0.2cm}$0.35 < (V-K)_0 < 1.15$ (4954 stars):
\begin{eqnarray}
\theta_{\rm eff}&=&0.5961+0.1567(V-K)_0+0.0309(V-K)_0^2+\nonumber \\
&&\!\!\!\!\!\!\!+\;\;0.009[m/H]+0.0022[m/H]^2\nonumber\\
&&\!\!\!\!\!\!\!+\;\;0.0021(V-K)_0[m/H] -0.0067 \log g\nonumber \\
\sigma_{\theta_{\rm eff}}&=& 0.0028
\label{vk-tef1}
\end{eqnarray}
\item \hspace{0.2cm}$1.15 \leq (V-K)_0 < 3.0$ (5820 stars):
\begin{eqnarray}
\theta_{\rm eff}&=&0.5135+0.2687(V-K)_0-0.0174(V-K)_0^2+\nonumber \\
&&\!\!\!\!\!\!\!+\;\;0.0298[m/H]-0.0009[m/H]^2 	\nonumber \\
&&\!\!\!\!\!\!\!-\;\;0.0184(V-K)_0[m/H] -0.0028 \log g\nonumber \\
\sigma_{\theta_{\rm eff}}&=&0.0026
\label{vk-tef2}
\end{eqnarray}
\end{itemize}
\noindent where $\theta_{\rm eff}={5040 \over T_{\rm eff}}$. The standard
deviation of Eqs. (\ref{vk-tef1}) and (\ref{vk-tef2}) is about 20 K and 25
K, respectively. As shown in Fig. \ref {resTempVK}, there is no residual
trend as a function of $(V-K)_0$, \feh or \logG. 
Equation (\ref{vk-tef1}) is aplicable in the range 
$3.25 \lesssim \log g \lesssim 4.75$ and eq. \ref{vk-tef2} in
the range $3.75 \lesssim \log g \lesssim 4.75$. Furthermore
the calibrations are valid in the ranges of 
colours and metallicities of the sample:
\begin{eqnarray}
-3.0 < [m/H] < -1.5 &\mbox{for}& 1.0 < (V-K)_0 < 2.9 \nonumber \\
-1.5 \leq [m/H] < -0.5 &\mbox{for}& 0.5 < (V-K)_0 < 2.9 \nonumber \\
-0.5 \leq [m/H] < 0.0 &\mbox{for}& 0.4 < (V-K)_0 < 3.0 \nonumber \\
 0.5 \leq [m/H] < 0.5 &\mbox{for}& 0.35 < (V-K)_0 < 2.8 
\label{ExpVal}
\end{eqnarray}

While $(V-K)_0$ is an observational quantity and \feh can be obtained from
photometric and/or spectroscopic measurements, a good determination of
\logg is usually unavailable for the most of the stars. This could
severely restrict the applicability of the above calibrations. However,
some photometric indexes, as the Str\"omgren $\delta c_1$ (see
\citet{crawford75a} or \citet{olsen88}), are good surface gravity
indicators and, if available, can help to estimate \logg. On the other
hand, catalogues of spectroscopic metallicities usually provide an
estimation of the surface gravity. Finally, a crude estimation of
\logg can be get from MK classification. The error in effective temperature
caused by an error in \logg will be:
\begin{equation}
\Delta T_{\rm eff} = \frac{a}{5040} T_{\rm eff}^2 \Delta \log g
\end{equation}
\noindent where $a$ is the coefficient of the \logg terms in Eqs.
(\ref{vk-tef1}) and (\ref{vk-tef2}). In the worst case, that of the hotter
stars, $\Delta T_{\rm eff} = 85~\Delta \log g$. Thus, even if the uncertainty
in \logg is as much as 0.5 dex, the error induced in \tefts is just 40 K.

\begin{figure}
\begin{center}
\resizebox{\hsize}{!}{\includegraphics{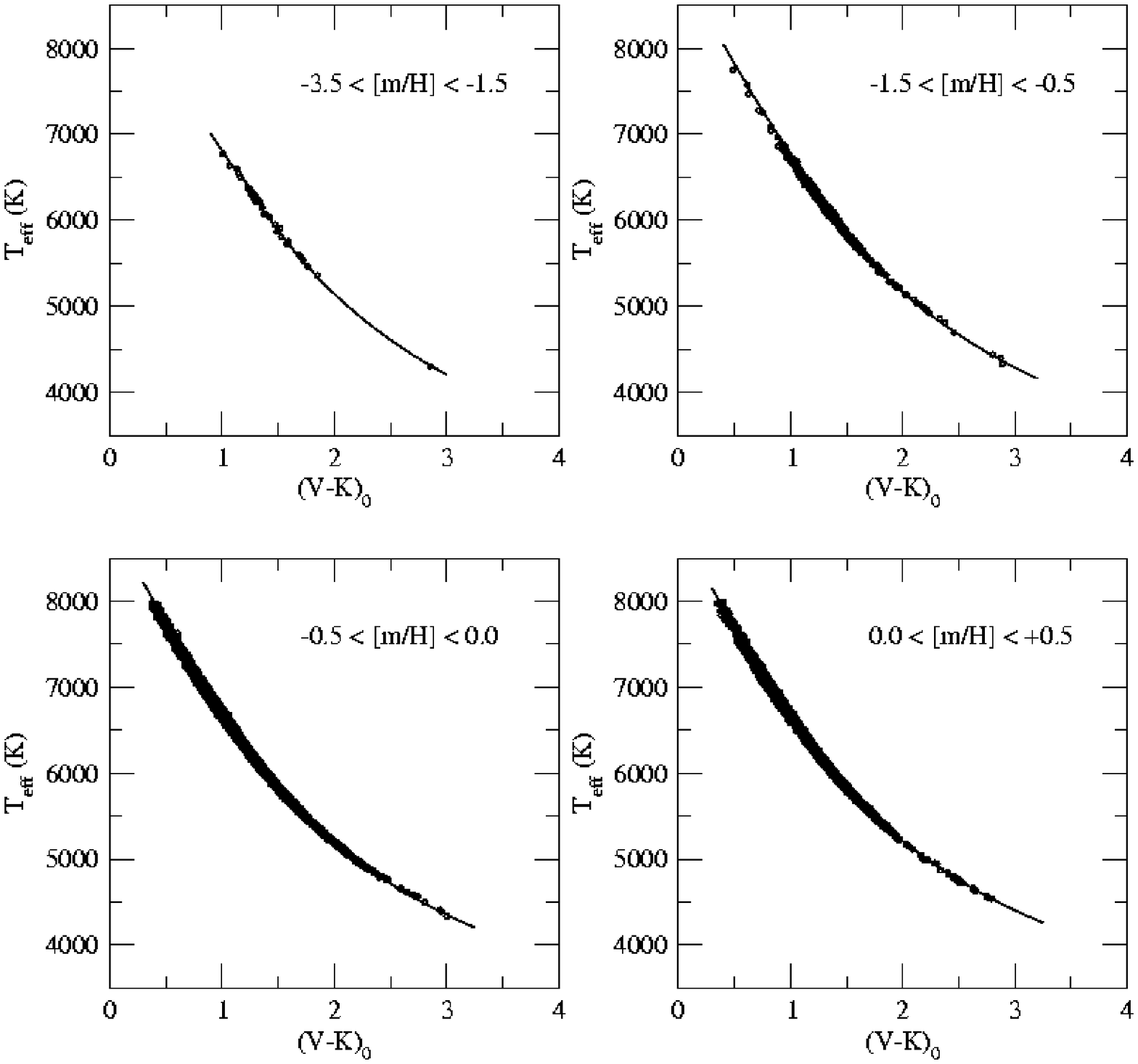}}
\end{center}
\caption{\tefts-$(V-K)_0$ fits for four groups of stars with different 
metallicities. The empirical relationships correspond to \logg=4.5 and
\feh=$-$2.0, $-$1.0, $-$0.25 and $+$0.25.}
\label{fitTeff}
\end{figure}

\begin{figure}
\begin{center}
\resizebox{\hsize}{!}{\includegraphics{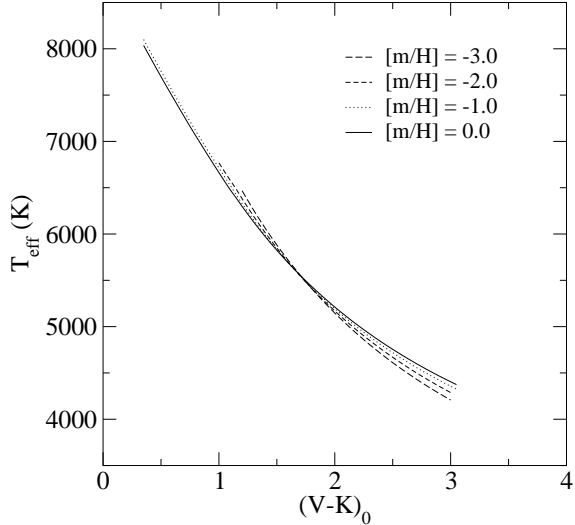}}
\end{center}
\caption{\tefts-$(V-K)_0$ relationships for \logg=4.5 and four different 
metallicities.}
\label{cTeff}
\end{figure}

The fits for four different metallicities and \logg = 4.5 together
with the stellar sample are shown in Fig. \ref{fitTeff}. Figure
\ref{cTeff} shows the empirical \tefts-$(V-K)_0$ relationships as a
function of the metallicity.

\subsection{Bolometric correction calibration}
\label{bolCorr}

Since the SEDF method provides both effective temperature and angular
semi-diameter, it also naturally allows for the determination of the
bolometric correction in a specific band. From this, if the distance is
known, one can compute the luminosity of the star. The bolometric
correction in a given band is defined as the difference between the
bolometric magnitude and the magnitude in that band:
\begin{equation}
BC_i = M_{\rm bol} - M_i = m_{\rm bol} - m_i
\label{Cbx}
\end{equation}
\noindent where $m_{\rm bol}$ and $m_i$ are assumed to be corrected of
interstellar reddening. $M_{\rm bol}$ can be easily expressed as a
function of the radius and effective temperature:
\begin{equation}
M_{bol} = -5 \log \frac{R}{\mbox{\rsun}} -10\log  
\frac{T_{\rm eff}}{T_{\rm eff\,\odot}}+ 4.74
\label{Mbol2}
\end{equation}
\noindent where \rsun=$6.95508\,10^8$ m and $T_{{\rm eff}\,\odot} = 5777$
K. For the Sun we adopt $V(\odot)=-26.75$ mag and $m_{\rm
bol}(\odot)=-26.83$ mag, and therefore $BC_V(\odot)=-0.08$ \citep{cox00}.

Using the definition of the absolute magnitude at a given band ($M_x = m_x
+ 5\log \pi +5$) and expressing the radius as function of the parallax
($\pi$) and the angular semi-diameter ($R=\theta/\pi$), we obtain the
following formula for the bolometric correction:
\begin{eqnarray}
BC_x & = & M_{\rm bol} - M_x = \nonumber \\
     & = & -5\log \left( {\cal{K}}\frac{\theta}{\mbox{\rsun}}\right) 
-10\log  \frac{T}{\mbox{\tsun}} -0.26 - m_x
\end{eqnarray}
\noindent where $\cal{K}$ is the factor corresponding to the
transformation of units. Once the bolometric correction for a band $i$ is
known, the bolometric correction for any band $j$ can be determined from:
\begin{equation}
BC_j  = (m_i-m_j) + BC_i
\label {CBxy}
\end{equation}
The error in the bolometric correction can be expressed as a function of the
uncertainties in \teft, $\theta$ (or $\cal{A}$) and $m_i$, as in Sect.
\ref{errors}:
\begin{eqnarray}
(\sigma_{BC_i})^2 & = & \left(\frac{5}{\ln 10}\frac{\sigma_{\theta}}{\theta} \right)^2 + \left(\frac{10}{\ln 10}\frac{\sigma_{T_{\rm eff}}}{T_{\rm eff}} \right)^2 + (\sigma_{m_i})^2 = \nonumber \\
 & = & (\sigma_{\cal{A}})^2 + \left(\frac{10}{\ln 10}\frac{\sigma_{T_{\rm eff}}}{T_{\rm eff}} \right)^2 + (\sigma_{m_i})^2
\end{eqnarray}

The procedure described here was used to compute the bolometric correction
in the $K$ (2MASS) band for the stars in our sample. In the same way
as for the effective temperature, we calibrated $BC$ as a function of
$(V-K)_0$, \feh and \logg with the following results:
\begin{itemize}
\item \hspace{0.2cm} $0.35 < (V-K)_0 < 1.15$ (4906 stars):
{\footnotesize
\begin{eqnarray}
BC_K&=&0.1275+0.9907(V-K)_0-0.0395(V-K)_0^2 +\nonumber \\
&&+0.0693[m/H]+0.0140[m/H]^2 \nonumber \\
&&+0.0120(V-K)_0[m/H]-0.0253\log g \nonumber \\
\sigma_{BC}&=&0.007~\mbox{mag} 
\label{vk-CBk1}
\end{eqnarray}
}
\item \hspace{0.2cm} $1.15 \leq (V-K)_0 < 3.0$ (5783 stars):
{\footnotesize
\begin{eqnarray}
BC_K&=&-0.1041+1.2600(V-K)_0-0.1570(V-K)_0^2 +\nonumber \\
&&+0.1460[m/H]+0.0010[m/H]^2 \nonumber \\
&&-0.0631(V-K)_0[m/H] -0.0079\log g\nonumber \\
\sigma_{BC}&=&0.005~\mbox{mag} 
\label{vk-CBk2}
\end{eqnarray}
}
\end{itemize}
The range of validity of these calibrations is the same as in the case of
the effective temperature. The bolometric correction in any band can be
obtained from $BC_K$ via Eq. (\ref{CBxy}).

\begin{figure}
\begin{center}
\resizebox{\hsize}{!}{\includegraphics{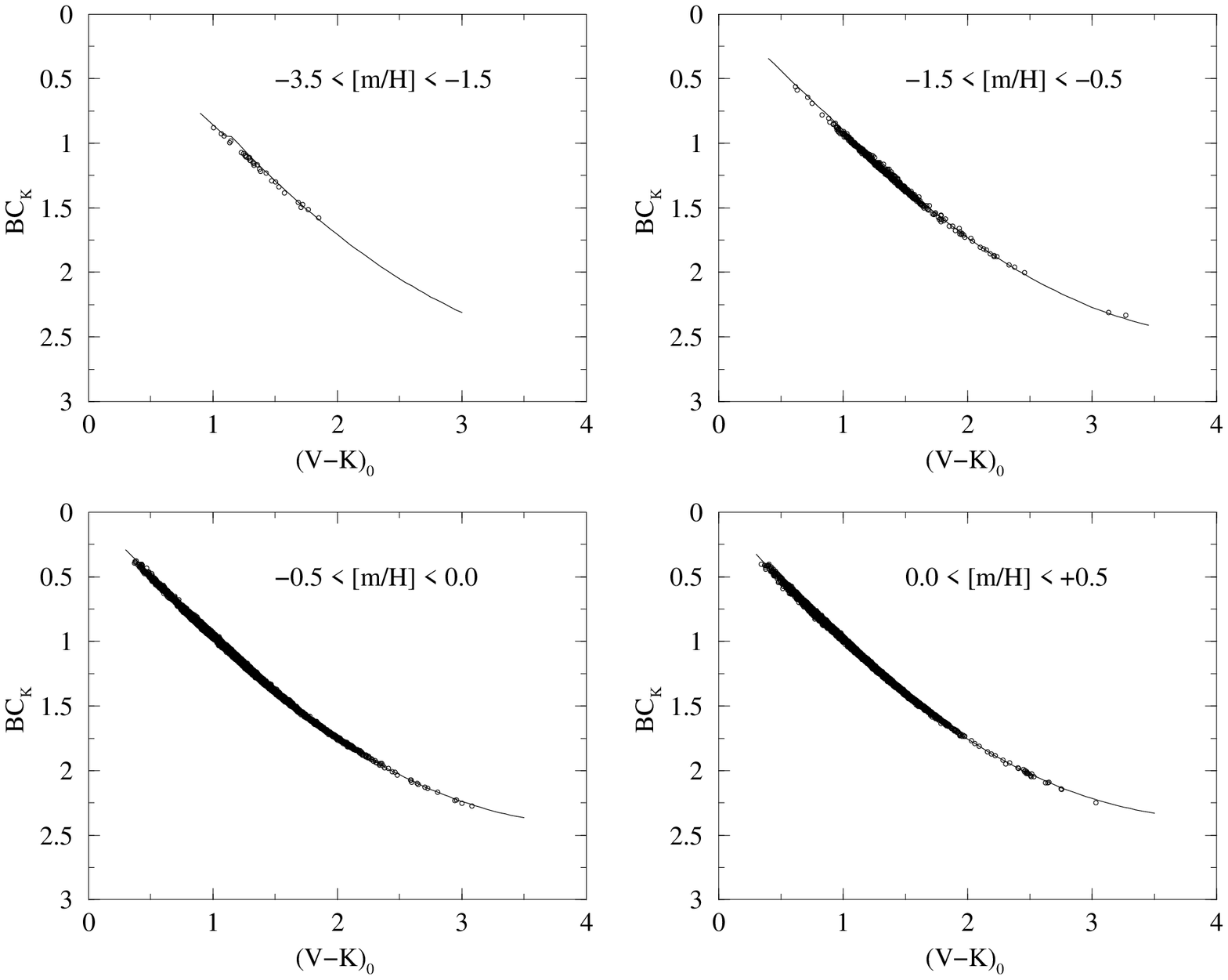}}
\end{center}
\caption{$BC_K-(V-K)_0$ fits for four groups of stars with different
metallicities. The empirical relationships correspond to \logg=4.5 and
\feh=$-$2.0, $-$1.0, $-$0.25 and $+$0.25.}
\label{fitBC}
\end{figure}

\begin{figure}
\begin{center}
\resizebox{\hsize}{!}{\includegraphics{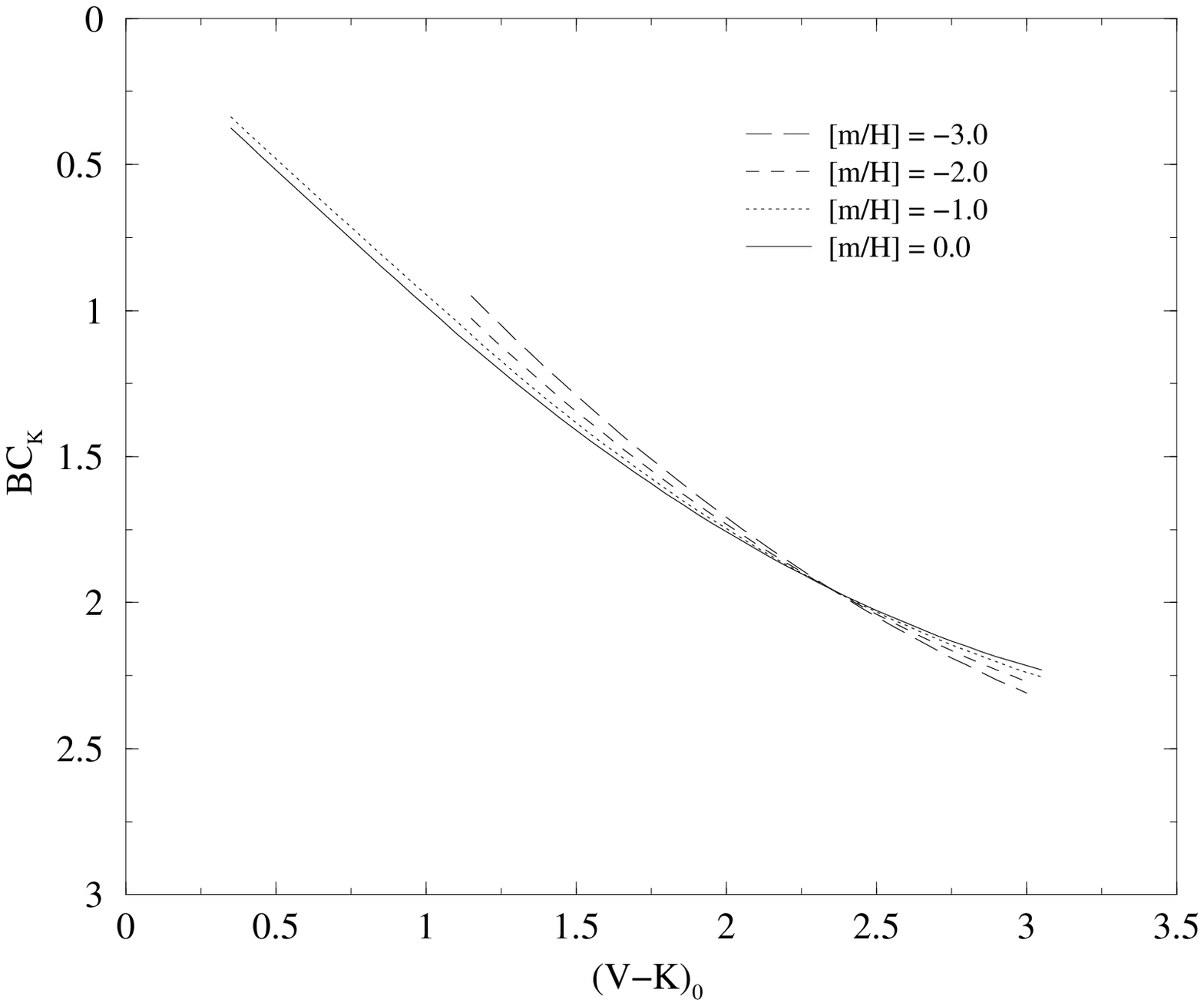}}
\end{center}
\caption{$BC_K-(V-K)_0$ relationships for \logg=4.5 and four different metallicities.}
\label{cBC}
\end{figure}

\begin{figure*}
\begin{center}
\includegraphics[width=15cm]{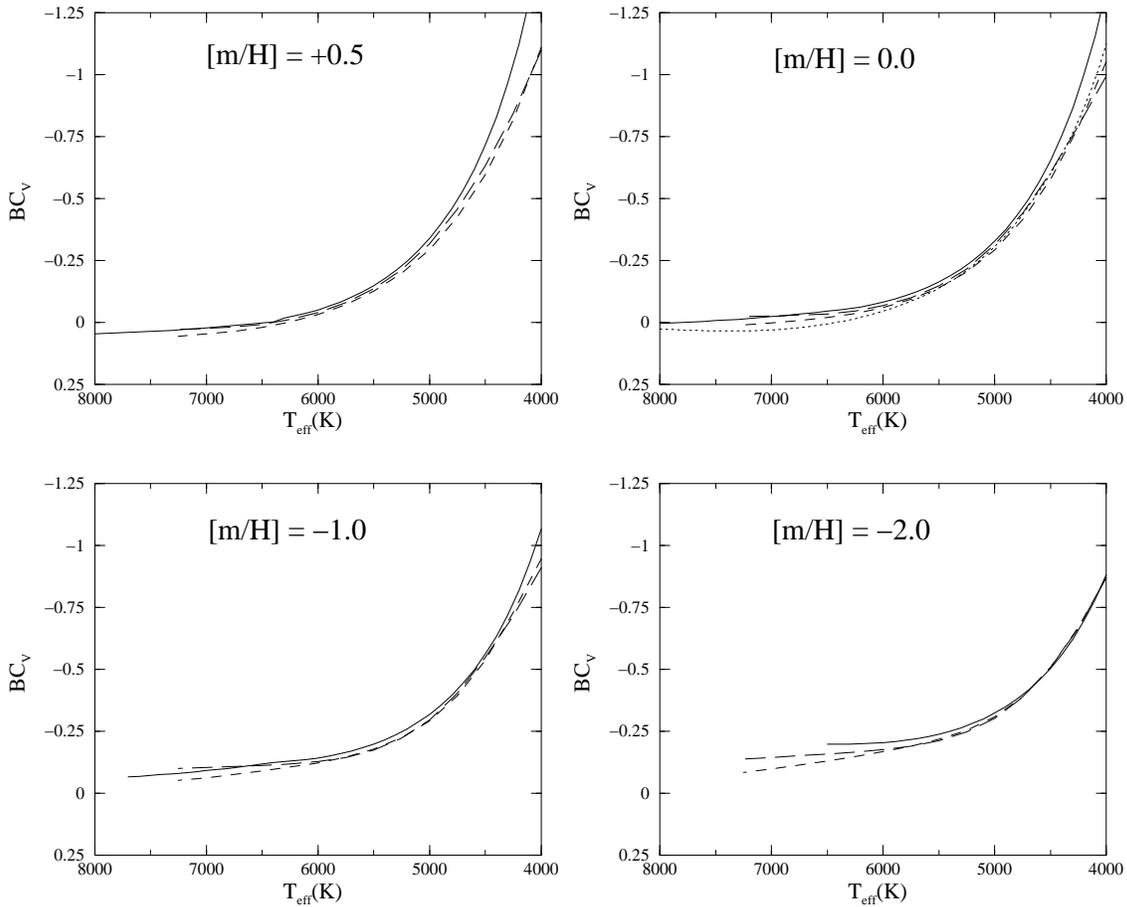}
\end{center}
\caption{Comparison between $BC_V$ values for \logg=4.5 given in Table \ref{BC_tef}
(solid line) and the values given by \citet {alonso95} for $\log g = 4$
(dashed line) and $\log g = 5$ (long-dashed line). In the panel
corresponding to $[M/H]=0.0$, the calibration of \citet{flower96} is also
shown (dotted line).}
\label{compCB}
\end{figure*}

Figure \ref{fitBC} shows the fits for four different metallicities,
together with the stars in the sample used to obtain the calibrations. The
$BC_K-(V-K)_0$ relationships as a function of the metallicity are shown in
Fig. \ref{cBC}. The calibration is tabulated in Table \ref{BC_tef} and
compared with the calibrations by \citet {alonso95} and \citet {flower96}
in Fig. \ref{compCB}, showing good agreement.

\begin{table*}
\caption {\sf Bolometric correction as a function of effective temperature
for \logg=4.5 and different metallicities (valid for dwarf and subdwarf stars). Tables for 
\logg=3.5 and \logg=4.0 are available in electronic form.}
\begin{center}
\begin{tabular}{ccccccccccc}
       \hline
         & \multicolumn{2}{c}{$[m/H] = +0.5$}    & \multicolumn{2}{c}{$[m/H] = 0.0$}   & \multicolumn{2}{c}{$[m/H] = -1.0$}   & \multicolumn{2}{c}{$[m/H] = -2.0$} & \multicolumn{2}{c}{$[m/H] = -3.0$}   \\
       \hline
   \teft &   $BC(V)$   &   $BC(K)$  &   $BC(V)$  &   $BC(K)$  &   $BC(V)$  &   $BC(K)$ &   $BC(V)$   &   $BC(K)$  &   $BC(V)$ &   $BC(K)$  \\
       \hline
       \hline
 4000  & $-$1.536 & 2.336 & $-$1.344 & 2.375 & $-$1.067 & 2.417 & ---      & ---   & ---      & ---   \\
 4100  & $-$1.310 & 2.324 & $-$1.156 & 2.348 & $-$0.933 & 2.371 & ---      & ---   & ---      & ---   \\
 4200  & $-$1.122 & 2.298 & $-$0.999 & 2.311 & $-$0.818 & 2.319 & ---      & ---   & ---      & ---   \\
 4300  & $-$0.964 & 2.262 & $-$0.865 & 2.266 & $-$0.719 & 2.263 & ---      & ---   & ---      & ---   \\
 4400  & $-$0.830 & 2.219 & $-$0.751 & 2.217 & $-$0.635 & 2.204 & $-$0.559 & 2.186 & $-$0.509 & 2.166 \\
 4500  & $-$0.716 & 2.171 & $-$0.653 & 2.164 & $-$0.562 & 2.144 & $-$0.504 & 2.121 & $-$0.468 & 2.097 \\
 4600  & $-$0.618 & 2.119 & $-$0.568 & 2.108 & $-$0.499 & 2.082 & $-$0.457 & 2.055 & $-$0.433 & 2.029 \\
 4700  & $-$0.532 & 2.064 & $-$0.495 & 2.050 & $-$0.444 & 2.020 & $-$0.416 & 1.990 & $-$0.403 & 1.962 \\
 4800  & $-$0.459 & 2.008 & $-$0.431 & 1.991 & $-$0.396 & 1.958 & $-$0.380 & 1.926 & $-$0.377 & 1.896 \\
 4900  & $-$0.395 & 1.950 & $-$0.376 & 1.932 & $-$0.355 & 1.896 & $-$0.350 & 1.862 & $-$0.356 & 1.831 \\
 5000  & $-$0.339 & 1.892 & $-$0.327 & 1.872 & $-$0.319 & 1.834 & $-$0.324 & 1.799 & $-$0.337 & 1.767 \\
 5100  & $-$0.290 & 1.833 & $-$0.285 & 1.813 & $-$0.288 & 1.773 & $-$0.301 & 1.737 & $-$0.321 & 1.704 \\
 5200  & $-$0.247 & 1.775 & $-$0.248 & 1.753 & $-$0.261 & 1.713 & $-$0.282 & 1.676 & $-$0.308 & 1.643 \\
 5300  & $-$0.209 & 1.717 & $-$0.216 & 1.695 & $-$0.237 & 1.653 & $-$0.265 & 1.616 & $-$0.298 & 1.582 \\
 5400  & $-$0.177 & 1.659 & $-$0.188 & 1.636 & $-$0.217 & 1.595 & $-$0.251 & 1.557 & $-$0.289 & 1.523 \\
 5500  & $-$0.148 & 1.601 & $-$0.163 & 1.579 & $-$0.199 & 1.537 & $-$0.239 & 1.499 & $-$0.282 & 1.466 \\
 5600  & $-$0.123 & 1.545 & $-$0.142 & 1.522 & $-$0.184 & 1.480 & $-$0.229 & 1.443 & $-$0.276 & 1.409 \\
 5700  & $-$0.101 & 1.489 & $-$0.123 & 1.467 & $-$0.171 & 1.425 & $-$0.221 & 1.387 & $-$0.272 & 1.354 \\
 5800  & $-$0.081 & 1.434 & $-$0.107 & 1.412 & $-$0.160 & 1.370 & $-$0.214 & 1.333 & $-$0.269 & 1.301 \\
 5900  & $-$0.065 & 1.380 & $-$0.093 & 1.358 & $-$0.151 & 1.317 & $-$0.209 & 1.280 & $-$0.267 & 1.248 \\
 6000  & $-$0.050 & 1.327 & $-$0.081 & 1.305 & $-$0.143 & 1.265 & $-$0.205 & 1.228 & $-$0.266 & 1.197 \\
 6100  & $-$0.038 & 1.275 & $-$0.071 & 1.253 & $-$0.137 & 1.213 & $-$0.202 & 1.178 & $-$0.266 & 1.147 \\
 6200  & $-$0.027 & 1.224 & $-$0.062 & 1.202 & $-$0.132 & 1.163 & $-$0.200 & 1.128 & $-$0.267 & 1.098 \\
 6300  & $-$0.018 & 1.173 & $-$0.055 & 1.152 & $-$0.128 & 1.114 & $-$0.199 & 1.080 & $-$0.268 & 1.050 \\
 6400  & $-$0.002 & 1.123 & $-$0.049 & 1.104 & $-$0.125 & 1.066 & $-$0.199 & 1.032 & $-$0.270 & 1.003 \\
 6500  &    0.003 & 1.074 & $-$0.045 & 1.054 & $-$0.118 & 1.022 & $-$0.199 & 0.986 & $-$0.272 & 0.957 \\
 6600  &    0.007 & 1.026 & $-$0.040 & 1.007 & $-$0.112 & 0.975 & ---      & ---   & ---      & ---   \\
 6700  &    0.011 & 0.978 & $-$0.035 & 0.959 & $-$0.106 & 0.929 & ---      & ---   & ---      & ---   \\
 6800  &    0.015 & 0.931 & $-$0.031 & 0.913 & $-$0.101 & 0.883 & ---      & ---   & ---      & ---   \\
 6900  &    0.019 & 0.884 & $-$0.026 & 0.866 & $-$0.096 & 0.838 & ---      & ---   & ---      & ---   \\
 7000  &    0.023 & 0.837 & $-$0.023 & 0.821 & $-$0.092 & 0.793 & ---      & ---   & ---      & ---   \\
 7100  &    0.026 & 0.792 & $-$0.019 & 0.775 & $-$0.087 & 0.748 & ---      & ---   & ---      & ---   \\
 7200  &    0.029 & 0.746 & $-$0.016 & 0.730 & $-$0.083 & 0.704 & ---      & ---   & ---      & ---   \\
 7300  &    0.032 & 0.701 & $-$0.012 & 0.686 & $-$0.079 & 0.660 & ---      & ---   & ---      & ---   \\
 7400  &    0.034 & 0.656 & $-$0.009 & 0.642 & $-$0.076 & 0.617 & ---      & ---   & ---      & ---   \\
 7500  &    0.037 & 0.612 & $-$0.007 & 0.598 & $-$0.072 & 0.574 & ---      & ---   & ---      & ---   \\
 7600  &    0.039 & 0.568 & $-$0.004 & 0.555 & $-$0.069 & 0.532 & ---      & ---   & ---      & ---   \\
 7700  &    0.041 & 0.525 & $-$0.002 & 0.512 & $-$0.066 & 0.490 & ---      & ---   & ---      & ---   \\
 7800  &    0.043 & 0.481 &    0.000 & 0.469 & ---      & ---   & ---      & ---   & ---      & ---   \\
 7900  &    0.044 & 0.438 &    0.002 & 0.427 & ---      & ---   & ---      & ---   & ---      & ---   \\
 8000  &    0.046 & 0.396 &    0.004 & 0.385 & ---      & ---   & ---      & ---   & ---      & ---   \\
\hline
\end{tabular}
\end{center}
\label{BC_tef}
\end{table*}

\section{Discussion}
\label{disc}

The procedure described in this paper yields three basic stellar
parameters: the best-fitting effective temperature and angular
semi-diameter and, from them, the bolometric correction. If the distance
is known, $\theta$ can be transformed into the true stellar radius. The
accuracies of the parameters for the stars in our sample are 0.5--1.3\% in
\teft, 1.0--2.5\% in $\theta$ and 0.04--0.08 mag for the \bC.

Comparisons with other determinations described in Sect.
\ref{comparacions} show general good agreement, with differences below
0.5$\sigma$, except for \citet {alonso96a} and \citet{ramirez05a}, 
where the difference is about
0.8$\sigma$. The use of different atmosphere models and the intrinsic
nature of the methods (photometric for \citeauthor{edvardsson93},
\citeauthor{alonso96a},  \citeauthor{ramirez05a} and ours; 
spectroscopic for \citeauthor{santos03}
and \citeauthor{fuhrmann98}) can explain in part the small differences. In
the case of the IRFM, the main difference between the implementation of
both \citet{alonso96a} and \citet{ramirez05a}, and 
the SEDF method is the absolute flux calibration:
\citet{alonso94b} for the IRFM and \citet{cohen03b} for the SEDF. This,
together with the use of different versions of the ATLAS9 atmosphere
models, is probably the reason for the $\sim$60 K differential between both
implementations of the IRFM and our determination. 
For \citet{ramirez05a} there is a dependence of $\Delta T_{\rm eff}$ with
$[m/H]$ in such a way that the temperature difference  
(\citeauthor{ramirez05a}- SEDF) increase abruptly for $[m/H] \lesssim 2.0$.
In all the other cases, the temperature differences 
are not correlated with \FEH. 

The most important factor to explain the systematics among the effective
temperatures computed from different methods is the absolute flux
calibration affecting photometric determinations and inaccuracies of
model atmospheres (non-LTE effects, 3D effects,
treatment on convection ,...) affecting both photometric and spectroscopic
determinations. \citet {bohlin04} pointed out a probable 2\% overestimation
of the IR flux in the Vega model used by \citet{cohen03a}.  A 2\% shift in
absolute flux calibration is equivalent to a difference of about 40 K in
temperature and to a zero point offset in the synthetic photometry of
0.022 mag. Such value would be compatible with our magnitude zero points
in Sect. \ref{analegs}.

Beyond the internal errors, which in the case of the SEDF take into
account the uncertainty in the flux calibration and all other error
sources, the comparison with other methods shows that, at present, the
systematic errors involved in the determination of effective temperature
are of about 20--30~K, equivalent to the 2\% uncertainty in the IR fluxes
of Vega claimed by \citet {bohlin04} to be a realistic value.

\section{Conclusions} 
\label{conclu}

We have presented a method (called SEDF) to compute effective
temperatures, angular semi-diameters and bolometric corrections from \MASS
photometry. We have adopted an approach based on the fit of the observed
$VJHK$ magnitudes using synthetic photometry, and it yields accuracies
around 1\% in \teft, 2\% in $\theta$, and 0.05 mag in $BC$, in the
temperature range 4000--8000 K.  A zero point offset was added to the
synthetic photometry computed from the Kurucz atmosphere models to tie in
our temperature scale with the Sun's temperature through a sample of solar
analogues. From the application to a large sample of FGK Hipparcos
dwarfs and subdwarfs 
we provide parametric calibrations for both effective temperature and
bolometric correction as a function of $(V-K)_0$, $[m/H]$ and \logG. Note that
the method presented here has been selected as one of the main sources of
effective temperatures to characterize the primary and secondary targets
of the COROT space mission \citep {baglin00}. Also, it is being currently
implemented as one of the tools offered by the Spanish Virtual Observatory
\citep{solano05}.

The resulting temperatures have been compared with several photometric and
spectroscopic determinations. Although we obtained remarkably good
agreement, slight systematic differences with other semi-empirical
methods, such as the IRFM, are present. This is probably due to the
uncertainties in the absolute flux calibration used by different
techniques. It is possible that, in spite of the great effort carried out
by \citet{cohen03a} and others to construct a consistent absolute flux
calibration in both the optical and the IR regions, some problems still
remain, which introduce small systematic effects in the temperatures.  
However, these effects seem to be as small as 20--30 K and could be
explained through uncertainties in the IR fluxes of about 2\%. In
conclusion, the results presented here strongly suggest that, given
the small differences found between methods, the effective temperature
scale of FGK stars (4000--8000 K) is currently established with a net
accuracy better than 0.5--1.0\%.

\begin{acknowledgements}
We are grateful to Dr. Angel Alonso for the suggestion of using solar
analogues to calibrate our method. We thank P. Nissen and the
anonymous referee for their remarks that helped to improve the paper. 
We also acknowledge support from the
Spanish MCyT through grant AyA2003-07736.  I.~R. acknowledges support from
the Spanish Ministerio de Ciencia y Tecnolog\'{\i}a through a Ram\'on y
Cajal fellowship. This publication makes use of data products from the Two
Micron All Sky Survey, which is a joint project of the University of
Massachusetts and the Infrared Processing and Analysis Center/California
Institute of Technology, funded by the National Aeronautics and Space
Administration and the National Science Foundation.
\end{acknowledgements}

\begin{table*}[!t]
\centering
\caption[]{Effective temperatures, angular semi-diameters, radii and 
bolometric corrections in both $V$ and $K$ bands, with their 
uncertainties, for 10999 stars as resulting from the algorithm presented 
in this paper.}
\label{tabTR}
\scriptsize
\begin{tabular}{rcccccrccccc}
\hline
\hline
\multicolumn{1}{c}{HD} &
$T_{\rm eff}$ &
$\theta$ &
$R$ &
$BC(V)$ &
$BC(K)$ &
\multicolumn{1}{c}{HIP} &
$T_{\rm eff}$ &
$\theta$ &
$R$ &
$BC(V)$ &
$BC(K)$ \\
 &
(K) & 
(mas) &
(R$_{\odot}$) &  
 &
 &
 &
(K) & 
(mas) & 
(R$_{\odot}$) &
& \\
\hline
 68988 & $\!\!\!$5911$\pm$47  & $\!\!\!$0.094$\pm$0.001  & $\!\!\!$1.183$\pm$0.044 &$-$0.068$\pm$0.019& 1.401$\pm$0.024  &145675                   & $\!\!\!$5357$\pm$32  & $\!\!\!$0.251$\pm$0.003  & $\!\!\!$0.978$\pm$0.013&$-$0.186$\pm$0.017& 1.798$\pm$0.020\\
\hline
\label{tabsample}
\end{tabular}
\end{table*}

\bibliographystyle{aa}
\bibliography{tesis}

\end{document}